\documentclass[twocolumn]{aastex61}

\usepackage{graphicx, subfigure, float, amsmath, natbib, epstopdf, grffile}

\newcommand{\Kepler}{\textit{Kepler }}
\renewcommand{\th}[1]{$#1^{\mathrm{th}}$}

\newcommand{\change}[1]{#1}

\shortauthors{Lurie et al.}

\begin{document}

%%%%%%%%%%%%%%%%%%%%%%%%%%%%%%%%%%%%%%%%%%%%%%%%%%%%%%%%%%%%%%%%%%%%%%
%%%%%%%%%%%%%%%%%%%%%%% TITLE PAGE %%%%%%%%%%%%%%%%%%%%%%%%%%%%%%%%%%%
%%%%%%%%%%%%%%%%%%%%%%%%%%%%%%%%%%%%%%%%%%%%%%%%%%%%%%%%%%%%%%%%%%%%%%

\title{Tidal Synchronization and Differential Rotation of \textit{Kepler} Eclipsing Binaries}

\correspondingauthor{John C. Lurie}
\email{lurie@uw.edu}

\author[0000-0002-8114-0835]{John C. Lurie}
\affiliation{Department of Astronomy\\
University of Washington \\
Seattle, WA 98195}

\author{Karl Vyhmeister}
\affiliation{Department of Astronomy\\
California Institute of Technology\\
Pasadena, CA 91125}

\author{Suzanne L. Hawley}
\affiliation{Department of Astronomy\\
University of Washington \\
Seattle, WA 98195}

\author{Jamel Adilia}
\affiliation{Department of Astronomy\\
University of Washington \\
Seattle, WA 98195}

\author{Andrea Chen}
\affiliation{Department of Astronomy\\
University of Washington \\
Seattle, WA 98195}

\author{James R. A. Davenport}
\affiliation{Department of Physics and Astronomy\\
Western Washington University \\
Bellingham, WA 98225}

\author{Mario Juri\'{c}}
\affiliation{Department of Astronomy\\
University of Washington \\
Seattle, WA 98195}

\author{Michael Puig-Holzman}
\affiliation{Department of Astronomy\\
University of Washington \\
Seattle, WA 98195}

\author{Kolby L. Weisenburger}
\affiliation{Department of Astronomy\\
University of Washington \\
Seattle, WA 98195}

\begin{abstract}

Few observational constraints exist for the tidal synchronization rate of late-type stars, despite its fundamental role in binary evolution. We visually inspected the light curves of 2278 eclipsing binaries (EBs) from the \Kepler Eclipsing Binary Catalog to identify those with starspot modulations, as well as other types of out-of-eclipse variability. We report rotation periods for 816  EBs with starspot modulations, and find that 79\% of EBs with orbital periods less than ten days are synchronized. However, a population of short period EBs exists with rotation periods typically 13\% slower than synchronous, which we attribute to the differential rotation of high latitude starspots. At 10 days, there is a transition from predominantly circular, synchronized EBs to predominantly eccentric, pseudosynchronized EBs. This transition period is in good agreement with the predicted and observed circularization period for Milky Way field binaries. At orbital periods greater than about 30 days, the amount of tidal synchronization decreases. We also report 12 previously unidentified candidate $\delta$ Scuti and $\gamma$ Doradus pulsators, as well as a candidate RS CVn system with an evolved primary that exhibits starspot occultations. For short period contact binaries, \change{we observe a period-color relation, and compare it to previous studies}. As a whole, these results represent the largest homogeneous study of tidal synchronization of late-type stars.

\end{abstract}


\section{Introduction}
\label{sec:intro}

At least half of star systems are binaries \citep{Duchene2013}, and many binaries are close enough that they will tidally interact. The evolution of stars in tidally interacting binaries is fundamentally different than for isolated stars. A tidally interacting system generally tends toward a state of equilibrium, where the orbit is circular, and the stellar rotation is coplanar and synchronized with the orbit \citep{Hut1980}. Tidal interaction can also lead to mass transfer and related phenomena including cataclysmic variables \citep{Warner2003}, supernovae \citep{Langer2012}, and degenerate object mergers \citep{Postnov2014}. Furthermore, tidal interaction can be used to probe the internal structure of stars \citep{Ogilvie2014}. Given the ubiquity of binaries and the importance of tidal interaction, observational constraints in this area are crucial to understanding stellar populations as a whole.

While numerous observational studies have focused on tidal circularization (e.g., \citealt{Koch1981,Duquennoy1991,Meibom2005,VanEylen2016}), progress on tidal synchronization has been limited by three major factors. First, stellar rotation rates are generally more difficult to measure than orbital periods. Second, most studies of synchronization have measured rotational velocities from line broadening. Conversion from rotational velocities to periods depends on the stellar radius and inclination, which may be uncertain. Third, and perhaps most importantly, most synchronization studies have focused on early-type stars with radiative envelopes (e.g., \citealt{Levato1974,Giuricin1984a,Abt2004,Khaliullin2010}). Only a few studies have focused on late-type stars with convective envelopes (\citealt{Giuricin1984b,Claret1995,Meibom2006,Marilli2007}), where the tidal dissipation mechanism is likely different than in radiative envelopes \citep{Zahn1977,Ogilvie2014}.

The \Kepler mission offers to greatly expand the number of rotation period measurements of tidally interacting binaries with convective envelopes, because of its unmatched ability to observe a large sample of eclipsing binaries (EBs) and to measure their rotation periods directly from starspot modulations. The Kepler Eclipsing Binary Catalog (KEBC\footnote{\url{http://keplerebs.villanova.edu}}, \citealt{Prsa2011,Slawson2011,Kirk2016}) contains over 2,800 candidate EBs observed during the original \Kepler mission. \Kepler has also revolutionized the study of the stellar rotation period distribution, with tens of thousands of rotation periods measured to date for single stars (e.g. \citealt{Harrison2012, McQuillan2014, Meibom2011, Nielsen2013, Reinhold2015}). 

While most \Kepler rotation studies have excluded stars with known stellar and substellar companions, \citet{Walkowicz2013} reported rotation periods for 950 exoplanet candidate (\Kepler Object of Interest) host stars. This study incidentally measured rotation periods for EBs that were misidentified as transiting exoplanets. 116 systems in that study are confirmed EBs in the \Kepler false positive list (\citealt{Bryson2015}), of which 48 have rotation periods within 25\% of the orbital period, suggesting that synchronization is occurring. However, rotation periods remain unmeasured for the vast majority of the KEBC.

Here, we systematically measure rotation periods for the KEBC, which allows us to investigate the dependence of tidal synchronization on several key orbital and stellar parameters. In the traditional paradigm, tidal energy is dissipated by convective turbulence in convective regions, and by radiative diffusion in radiative regions \citep{Zahn1977}. These two processes proceed at different rates, and the rate of tidal evolution for a given star depends on the locations and relative thicknesses of its convective and radiative regions. The rate of tidal interaction also depends on the mass ratio, with the rate increasing for more equal mass binaries. Also, tidal forces are stronger at smaller separations, so shorter period EBs should be more synchronized. However, a state of true synchronization is impossible in eccentric binaries. Instead, the binary approaches ``pseudosynchronization'', where the rotational angular velocity synchronizes to the orbital angular velocity at periastron, where the tidal forces are the strongest \citep{Hut1981}. Thus, mass, mass ratio, orbital period, and eccentricity are all important parameters to investigate.

\change{An unexpected result of our investigation is a population of EBs that are rotating typically 13\% slower than synchronous. After ruling out instrumental and numerical causes, differential rotation is the most likely physical explanation. Differential rotation is important to binary evolution in its own right, as it influences magnetic braking through surface activity and the magnetic dynamo \citep{Schatzman1962}. \citet{Reinhold2013} and \citet[][hereafter RG15]{Reinhold2015} presented differential rotation measurements for thousands of single \Kepler stars, and examined trends with effective temperature and rotation period. Using a similar technique, we measure differential rotation for the EBs, and demonstrate how differential rotation explains the subsynchronous population of EBs.}

The remainder of the paper is organized as follows. In $\S$\ref{sec:data}, we describe the KEBC and the \Kepler light curves. In $\S$\ref{sec:analysis}, we classify the EB light curves and measure rotation periods for EBs with starspot modulations. In $\S$\ref{sec:results}, we examine the dependence of tidal synchronization on orbital period, eccentricity, stellar mass, and mass ratio, while in $\S$\ref{sec:diff rot} we focus on differential rotation. We present additional results in $\S$\ref{sec:additional results}, and conclude in $\S$\ref{sec:summary}.

\section{Data}
\label{sec:data}

\subsection{The \Kepler Eclipsing Binary Catalog}
We began with the 2863 targets in the KEBC downloaded on 2017 March 24. The KEBC includes orbital periods, ephemerides, and primary and secondary (when detected) eclipse depths, widths and phase separations.  There are some uncertainties in the KEBC that are relevant to our analysis. A circular EB with nearly equal primary and secondary eclipse depths may be mistaken for an EB with only a primary eclipse at half the given period. Some systems with small eclipse depths may be transiting exoplanets or brown dwarfs, although most have been removed by the KEBC and \Kepler mission teams. Although substellar companions are not the focus of this work, we include them in our analysis for completeness. In $\S$\ref{subsubsec:outliers}, we use rotation period measurements to identify when the above cases occur.

We excluded the following targets from our sample. There are 11 systems with eclipses at multiple periods (nine with two periods and two with three periods) due to the ambiguity of assigning orbital periods to a measured rotation period. There are 406 targets flagged as uncertain in the KEBC, most of which are contact binaries or ellipsoidal variables and would not have been analyzed in any case. There are 168 targets flagged as heartbeat stars \citep{Kumar1995,Thompson2012}, which we excluded due to the complex light curves and extreme dynamics of these systems. After these exclusions, there were 2278 EBs remaining that we analyzed.

\subsection{Eccentricity}
\label{subsec:eccentricities}

Constraints can be placed on the orbital eccentricity from the timing and relative durations of primary and secondary eclipses\footnote{The eccentricity may also be constrained using the duration differences between ingress and egress \citep{J.Barnes2007,R.Barnes2015}.}. These constraints are uncertain upon the argument of periastron $\omega$. In a circular orbit, the primary and secondary eclipses will be separated in phase by 0.5, and will have the same duration, regardless of $\omega$.

Using the timings of primary and secondary eclipses $t_{pri}$ and $t_{sec}$, $e\cos{\omega}$ can be approximated as
\begin{equation}
e \cos{\omega}  \approx \frac{\pi}{2} \left(\frac{| t_{sec} - t_{pri} |}{P_{orb}} - \frac{1}{2} \right)
\end{equation}
If $t_{pri} - t_{sec} = P_{orb}/2$, then $e\cos{\omega} = 0$. This corresponds to either a circular orbit, or an eccentric orbit with $\omega = 90^{\circ}$. If $t_{pri} - t_{sec} \ll P_{orb}$, then $e |\cos{\omega}|$ approaches a maximum value of 1, corresponding to a highly eccentric orbit.

From the durations of primary and secondary eclipses $d_{pri}$ and $d_{sec}$, $e\sin{\omega}$ can be approximated as
\begin{equation}
e\sin{\omega} \approx \frac{(d_{sec}/d_{pri} - 1)}{(d_{sec}/d_{pri} + 1)}
\end{equation}

An approximation of the eccentricity can then be determined from the combination of $e\cos{\omega}$ and $e\sin{\omega}$. Constraining the eccentricity in this way requires an EB with detected primary and secondary eclipses. This favors binaries with comparable surface temperatures and relatively small orbital separations. Of the 816 EBs in our rotation period catalog (see $\S$\ref{subsec:classification}), 484 have eccentricity constraints using this method.

We stress that these eccentricities should only be regarded as approximations for the purposes of studying bulk trends with eccentricity. The KEBC does not include uncertainties on the eclipse timings and durations, and therefore we cannot propagate the uncertainties in our calculations. A fuller treatment of the uncertainties would require intensive modeling that is beyond the scope of this work. Ultimately this is of little concern, as we are most interested in differences in synchronization between clearly circular and clearly eccentric systems, rather than the exact dependence on eccentricity.

\subsection{\Kepler Light Curves}

We analyzed \Kepler quarters 0-17 light curves from Data Release 25. We used the Simple Aperture Photometry (SAP) fluxes, detrended by the \Kepler mission pipeline Presearch Data Conditioning (PDC). Cadences were excluded if they had  \texttt{SAP\_QUALITY} flag values of 128, indicating that a cosmic ray was found and corrected in the optimal aperture, or 2048, indicating that an impulsive outlier was removed before detrending \citep{Thompson2016}. For each quarter, we subtracted and then divided by the median flux value. The resulting dimensionless relative flux values are useful for intercomparing EB light curves, and are necessary for the autocorrelation function method to measure rotation periods (see $\S$\ref{subsec:measure periods}).

The current PDC pipeline suppresses stellar variability at periods longer than approximately 20 days \citep{Gilliland2015}. There is a tradeoff between using undetrended SAP and detrended PDC light curves. By using the PDC light curves, we are more confident that the rotation periods we measure are not due to instrumental artifacts, but we may detect fewer slowly rotating stars. If we had used the undetrended SAP light curves, we may have found more slow rotators, but would be less sure that they were astrophysical in origin. Even without the pipeline suppression, slow rotators are intrinsically more difficult to detect, because the amplitude of their starspot modulations is lower \citep{McQuillan2014}. Given these limitations, our synchronization study is primarily focused on EBs with rotation periods less than 20 days.

\section{Classification and Rotation Period Analysis}
\label{sec:analysis}

Our analysis involved two steps. First, we visually inspected the light curves to classify EBs with starspot modulations, as well as other types of EBs. Next, we measured the rotation periods for EBs with starspot modulations.

\subsection{Light Curve Classification}
\label{subsec:classification}

\begin{figure*}[ht]
\centering
\includegraphics[height=9in, trim = 0cm 0cm 0cm 0cm, clip]{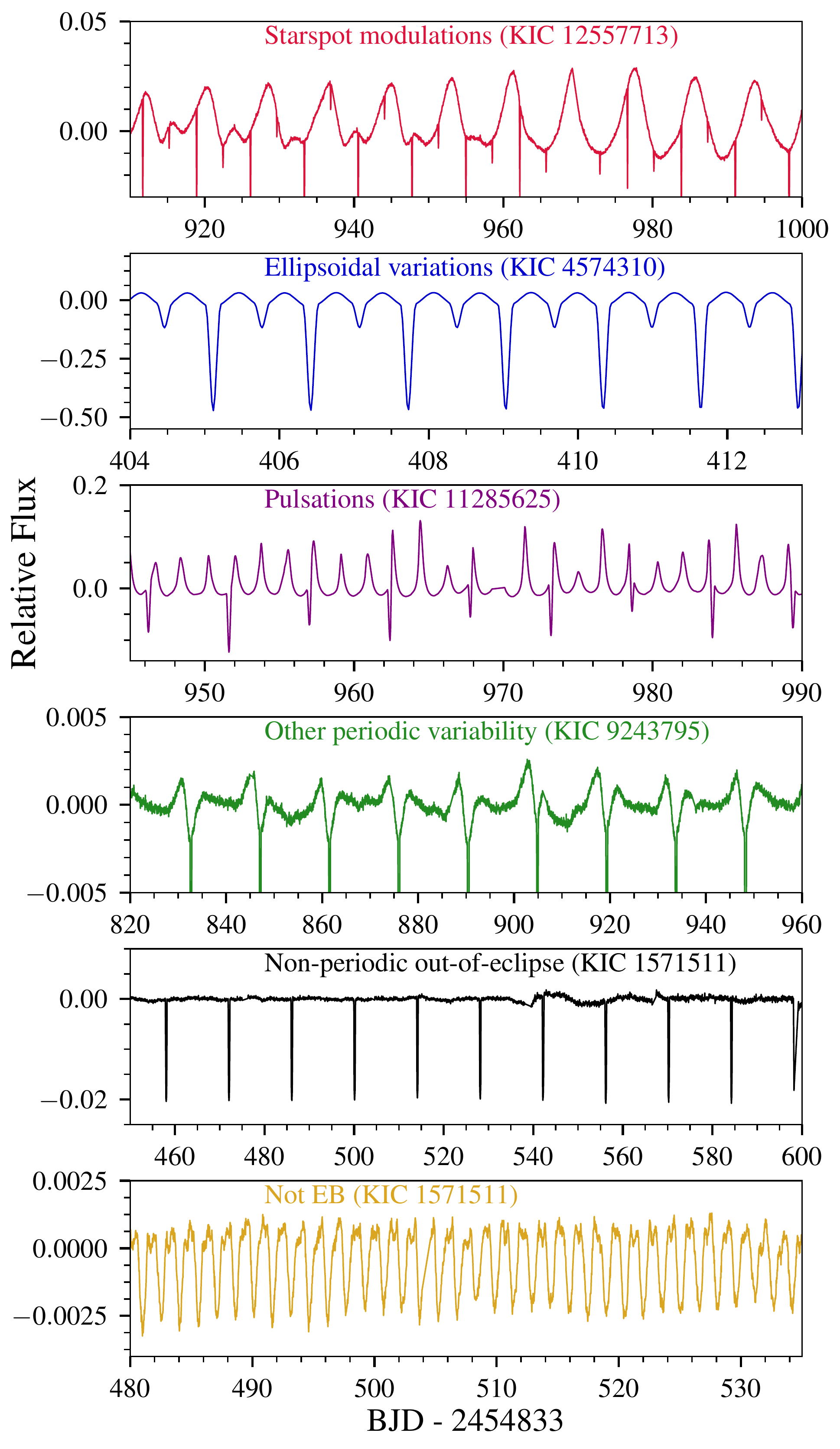}
\caption{Example light curves for the six classification types.}
\label{fig:class examples}
\end{figure*}

Light curves were divided into six categories based on the morphology of their out-of-eclipse variability. Examples are shown in Figure \ref{fig:class examples}.

% Figure 1: Classification examples

\begin{enumerate}

\item There are 816 EBs with starspot modulations (SP). These appear as roughly sinusoidal variations, and are due to periodic dips in brightness as spots (or spot groups) rotate into and out of view. The key feature of starspot modulation we used for classification is the phase and amplitude evolution of the modulations. An example of this evolution is shown in the top panel of Figure \ref{fig:class examples}. Between days 910 and 945, the out-of-eclipse variability has two humps. Between days 945 and 965, the smaller hump disappears, and the amplitude of the larger hump increases. This is due to the combination of differential rotation of the star, and the formation and dissipation of starspots. For a schematic of how differential rotation and spot evolution change the light curve appearance, see Figure 4 of \citet{Davenport2015}.

\item There are 779 EBs with ellipsoidal variations (EV). Ellipsoidal variations are due to the changing apparent cross section of the tidally distorted stars as they orbit each other. The stars have the largest cross sections at quadrature, resulting in two peaks in the light curve halfway between the primary and secondary eclipse. Unlike starspot modulations, ellipsoidal variations do not evolve over the 4 year observation baseline of \textit{Kepler}.

This category includes EBs with well-defined eclipses such as in the second panel of Figure \ref{fig:class examples}, and contact binaries without well-defined eclipses. Most EBs with ellipsoidal variations are likely circularized and synchronized due to the strong tidal forces at their small separations. However, ellipsoidal variations do not constitute a direct measurement of stellar rotation, and are not the focus of this work.

\item There are 27 EBs with $\delta$ Scuti and $\gamma$ Doradus pulsations (PU) and 21 with possible pulsations (PUX). An example is shown in the third panel of Figure \ref{fig:class examples}. In Table \ref{tab:pulsators}, we note 12 EBs that are not listed as pulsators in the KEBC or the literature.

\item There are 27 EBs with other periodic out-of-eclipse variability that is not due to one of the above phenomena (OT). Some of these may be previously unidentified heartbeat stars, such as the example shown in the fourth panel of Figure \ref{fig:class examples}.

\item There are 598 EBs without any clear periodic out-of-eclipse variability (NP), like that in the fifth panel of Figure \ref{fig:class examples}. Many of these have essentially flat out-of-eclipse light curves, or long term, smooth variations due to instrumental effects. Some EBs in this category have low level variability that may be due to starspots, but were too ambiguous to include in the starspot modulation category.

\item There are 10 targets where starspot modulations appear to have been mistaken for ellipsoidal variations, an example of which is shown in the bottom panel of Figure \ref{fig:class examples}. Due to the lack of clear eclipses, these targets may not be EBs. 

\end{enumerate}

% Table 1: Newly identified pulsators
\begin{deluxetable}{l}
\label{tab:pulsators}

\tabletypesize{\small}
\tablewidth{0pt}
\tablecaption{Previously Unidentified Pulsators}

\tablehead{
\colhead{Likely Pulsators (PU)}}

\startdata
KIC 10549576 \\
KIC 11724091 \\
KIC 11817750 \\
\hline  
\multicolumn{1}{c}{Candidate Pulsators (PUX)}\\
\hline  
KIC 5565486 \\
KIC 6063448 \\
KIC 6109688 \\
KIC 6145939 \\
KIC 6147122 \\
KIC 9552608 \\
KIC 11923819 \\
KIC 12106934 \\
KIC 12167361 \\
\enddata

%\tablecomments{}
% \tablenotetext{}{}
\end{deluxetable}

\subsection{Measurement of Rotation Periods}
\label{subsec:measure periods}

We measured rotation periods for the 816 EBs with starspot modulations using the following procedure. First, we linearly interpolated over eclipses, and then measured initial rotation periods using the autocorrelation function (ACF, see \citealt{McQuillan2013a}). The ACF is not very sensitive to multiple rotation period signals in the light curve that may originate from the two separate stars in an EB \citep{Rappaport2014}, or from differential rotation on one star (RG15). We therefore searched for multiple rotation periods using the Lomb-Scargle periodogram \citep{Lomb1976,Scargle1982}, with the ACF-based periods serving as a validation.
\subsubsection{Interpolation over Eclipses}

Eclipses are a source of contamination and were removed prior to measuring rotation periods. We linearly interpolated over windows around the eclipses that were equal to 1.5 times the eclipse widths listed in the KEBC. This larger window ensures that the eclipses are entirely removed. Interpolating over the eclipses does not adversely affect the rotation period measurements, because the EBs with starspot modulations typically have small eclipse widths; more than 83\% have total eclipse widths (primary plus secondary) less than 10\% of the total orbit.

% Figure 2: Rotation period measurement example
\begin{figure*}[ht]
\centering
\includegraphics[width=4in]{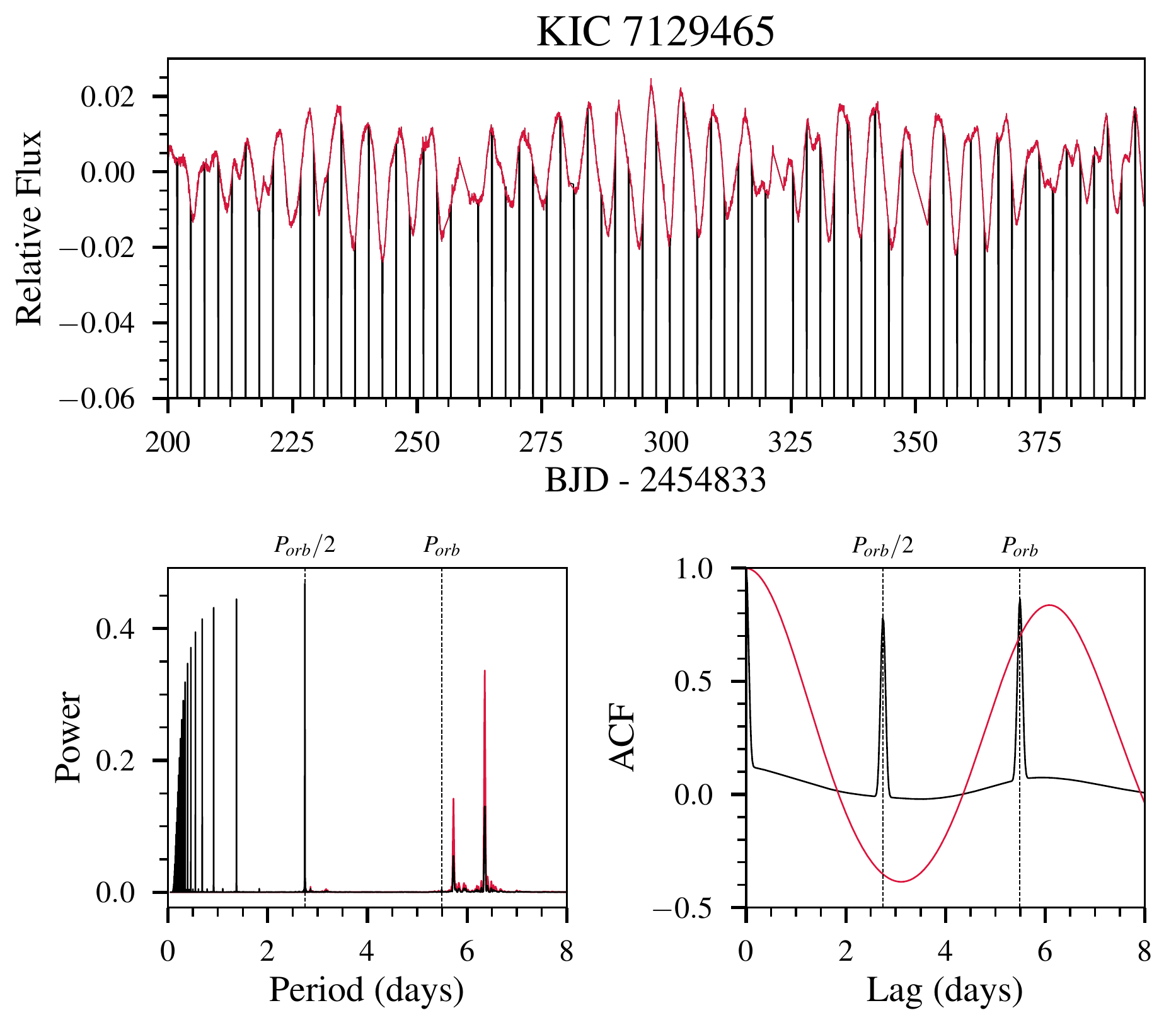}
\caption{Eclipse removal and rotation period measurements for the representative EB KIC 7129465. Top panel: A 200 day segment of the full 1460 day light curve. The out-of-eclipse light curve is plotted in red, while eclipses are plotted in black. The flux range has been truncated to focus on the out-of-eclipse variability.  Bottom left panel: Lomb-Scargle periodograms for the full (black), and out-of-eclipse (red) light curves. The black periodogram has been multiplied by a factor of five for clarity. Bottom right panel: Autocorrelation functions for the full (black) and out-of-eclipse (red) light curves.}
\label{fig:eclipse removal}
\end{figure*}

\subsubsection{Initial Periods from the Autocorrelation Function}

The ACF computes the self-similarity of a light curve at different time lags. Periodically varying light curves have a peak ACF value at the time lag corresponding to that period. We identify the peak in the ACF using the procedure of \citet{McQuillan2013a}. The ACF was first smoothed using a Gaussian filter with a kernel standard deviation of 18 time lags and window size of 56 time lags. In general, the first peak in the ACF is the highest, and corresponds to the stellar rotation period. However, if there are spots on opposite hemispheres, there will be a lower ACF peak at half of the rotation period. We manually corrected such instances, as well as cases where peaks at longer time lags were erroneously identified by the automated code.

Figure \ref{fig:eclipse removal} demonstrates the rotation period measurement for the 5.5 day orbital period EB KIC 7129465. There is a dramatic difference in the ACFs before and after eclipse removal. The black ACF (with eclipses) has sharp peaks at the half and full orbital periods due to the strong periodic signal of the eclipses. In contrast, the red ACF (without eclipses) has a wider peak at 6.1 days, somewhat longer than the orbital period. The shape of the red ACF is similar to those for single stars with starspot modulations \citep{McQuillan2013a}. This indicates that the eclipses have been successfully removed, and that the rotation period is longer than the orbital period, in this case.

% Figure 3: ACF relative peak height distributions
\begin{figure}[ht]
\centering
\includegraphics[width=3in, trim = 0cm 0cm 0cm 0cm, clip]{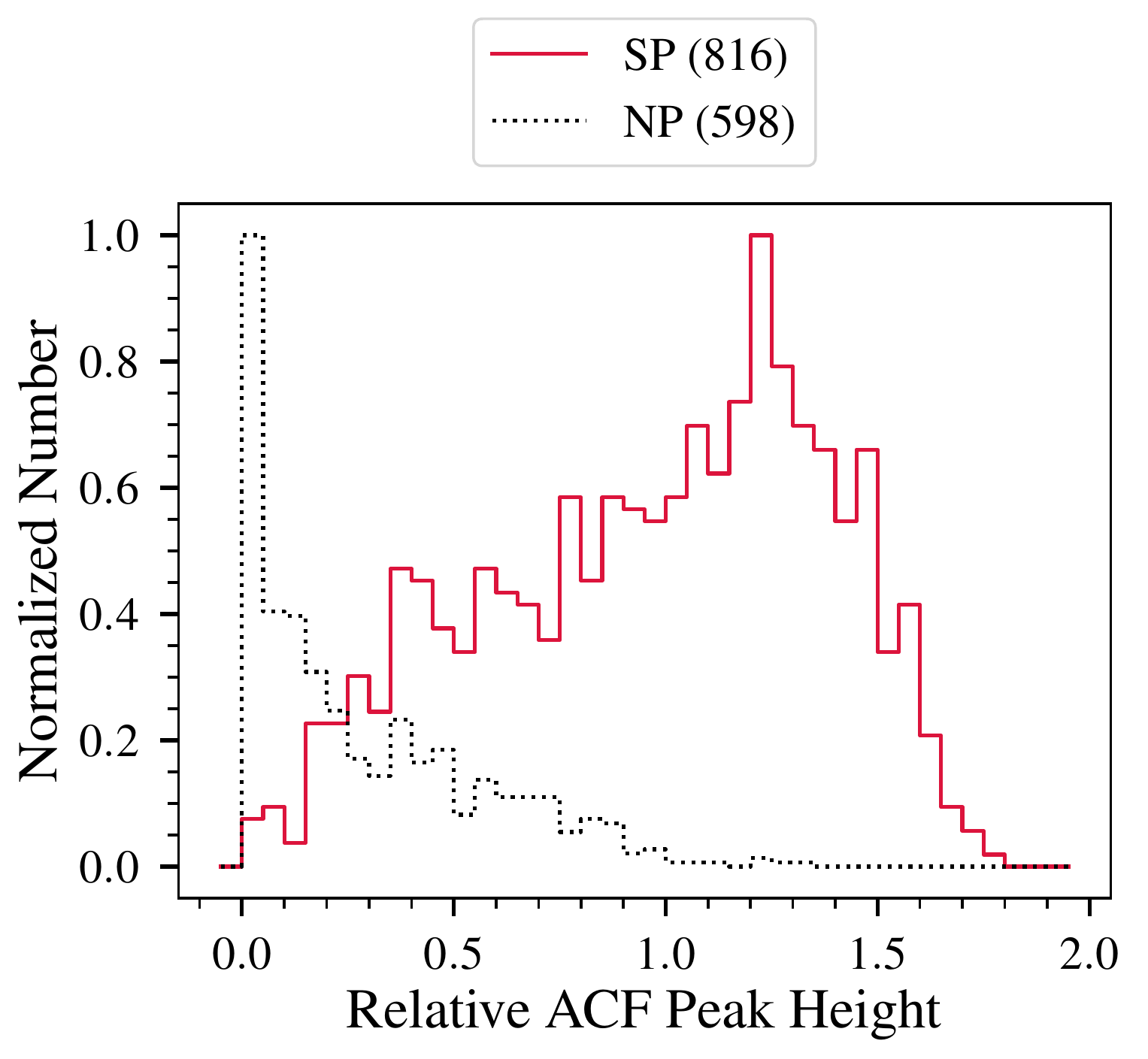}
\caption{Distributions of relative ACF peak heights for EBs with starspot modulations (SP, solid red line) and non-periodic out of eclipse variability (dotted black line). The histograms are normalized to their maximum values, and the number of EBs in each category are listed in parentheses. The likely starspot systems have the highest peaks, corresponding a strong periodic signal.}
\label{fig:acf peak heights}
\end{figure}

As further validation of our rotation periods, we compare the ACF peak heights of EBs with starspot modulations to the EBs without periodic out-of-eclipse variability. Following \citet{McQuillan2013a}, we define the peak height as the height of the ACF peak relative to the adjacent minima. Unlike the absolute height, the relative height is less susceptible to systematic effects in the light curve such as long term trends. The ACF has values between $-1$ and 1, so the relative peak height has values between 0 and 2.

Figure \ref{fig:acf peak heights} shows the distribution of ACF relative peak heights for EBs with starspot modulations and no periodic out-of-eclipse variability. The two distributions are clearly separated. 84\% of the starspot modulation category have peak heights greater than 0.5, compared to only 24\% for the non-periodic category. This provides validation that the EBs classified as having starspot modulations do exhibit significant out-of-eclipse periodicity.

\subsubsection{Multiple Rotation Periods from the Periodogram}
\label{subsubsec:multiple periods}

The bottom left panel of Figure \ref{fig:eclipse removal} shows periodograms for the full light curve (black) and after eclipses have been removed (red). The black periodogram has peaks at the half orbital period and lower harmonics, as is typical for EB periodograms. There are also two smaller peaks at 5.7 and 6.4 days, again somewhat longer than the orbital period. When the eclipses are removed (red periodogram), the orbital period harmonic peaks essentially disappear, and only the peaks at 5.7 and 6.4 days remain. Importantly, the locations of the peaks do not change, meaning that the removal of the eclipses cannot be responsible for the peaks. 

As discussed below, many of the EBs in our sample have two peaks in their periodograms after eclipses have been removed. This highlights the importance of using both the ACF and the periodogram. Had we only relied on the ACF, we would have missed information in the light curves. Meanwhile, the ACF provides validation that the peaks are due to starspot modulations.

We identified multiple rotation periods adapting the procedures of \citet{Rappaport2014} and RG15. This involved generating periodograms for each EB on a uniform frequency grid using the Python package \texttt{gatspy} \citep{VanderPlas2015,VanderPlas2016}. The grid had frequency bin widths of $5.7\times10^{-5}$ day$^{-1}$, which resulted in a very oversampled periodogram, as was desired. We then smoothed each periodogram using a Gaussian filter with a kernel standard deviation of 30 frequency bins and a window size of 120 bins. 

We searched for peaks in the smoothed periodogram in a period range from 2/3 of the ACF rotation period up to 200 days. Next, we identified the two highest significant peaks in the smoothed periodogram. Peaks were defined as significant if their heights were at least 30\% that of the highest peak. Lowering this threshold increases the possibility of finding multiple rotation periods, but also increases the possibility of finding spurious signals.

Next, we identified neighborhoods around the two peak groups in which we can search for subpeaks. The neighborhood is the frequency range between the local minima to the left and right of dominant peaks in the smoothed periodogram. Within each neighborhood, we applied the same threshold that significant subpeaks must be at least 30\% as high as the highest subpeak in the group. We then selected the subpeaks with the largest frequency separation.

% Figure 4: Multiple peak finding example
\begin{figure*}[ht!]
\centering
\includegraphics[width=5in, trim = 0cm 0cm 0cm 0cm, clip]{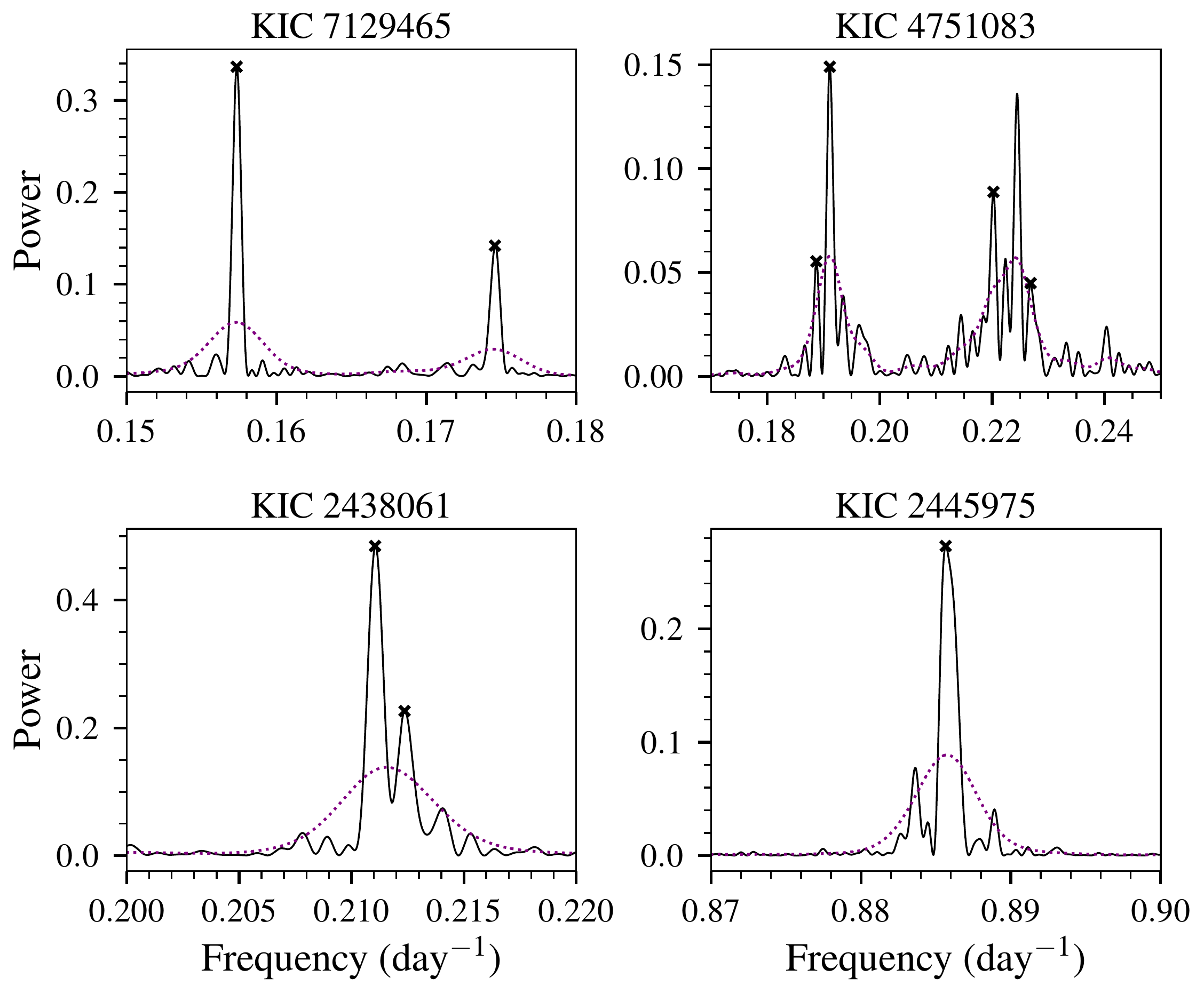}
\caption{Examples of the routine to find multiple rotation periods in the Lomb-Scargle periodogram. The solid black curve shows the oversampled periodogram, while the dashed purple curve shows the periodogram smoothed with a Gaussian filter. The black crosses indicate the significant subpeaks within each group.}
\label{fig:multipeak}
\end{figure*}

Figure \ref{fig:multipeak} demonstrates our multiple peak-finding algorithm for four representative cases. In the case of KIC 7129465 (top left panel), there are groups of peaks at 5.71 and 6.41 days (0.175 and 0.156 day$^{-1}$), as discussed above. However, there is only one significant subpeak in each group. In the case of KIC 4751083 (upper right), there are two well separated peak groups, with two significant subpeaks in each group. In the case of KIC 2438061 (lower left), there is only one significant group of peaks, and there are two significant subpeaks in that group. Finally, in the case of KIC 2445975 (lower right), there is only one significant subpeak, and hence no detection of multiple rotation periods.

We define a conservative rotation period limit of 45 days. Robustly measuring longer period signals is difficult due to instrumental systematics that differ between \Kepler quarters. Quarters are approximately 90 days long, and a cutoff of 45 days requires that we would see the rotation signal repeat twice in a single quarter. We do measure rotation periods longer than 45 days, but they should be treated with caution. This cutoff has a minimal effect on our synchronization analysis ($\S$\ref{sec:results}), which is primarily focused on the one to twenty day rotation period range.

\section{Tidal Synchronization}

In this section, we give a brief overview of our rotation period catalog, and the orbital period distribution of the different EB categories. We then use the catalog to investigate the dependence of tidal synchronization on orbital period, eccentricity, stellar mass, and mass ratio.

\label{sec:results}

\subsection{Rotation Period Catalog}

% Table 2: Rotation period catalog
\begin{deluxetable*}{lrlrrrrrrrrrrl}

\tabletypesize{\small}
\tablewidth{0pt}
\tablecaption{EB Classifications and Rotation Periods - Representative Subset}

\tablehead{
\colhead{KIC ID}&
\colhead{$P_{orb}$}&
\colhead{Class.} &
\colhead{$P_{ACF}$} &
\colhead{$h_{acf}$} &
\colhead{$P_{1,min}$} &
\colhead{$P_{1,max}$} &
\colhead{$P_{2,min}$} &
\colhead{$P_{2,max}$} &
\colhead{$h_{1,min}$} &
\colhead{$h_{1,max}$} &
\colhead{$h_{2,min}$} &
\colhead{$h_{2,max}$} &
\colhead{\change{Notes}}
}

\startdata
2997455 &  1.130 &    SP &  1.124 & 0.652 &   1.127 &   1.131 &         &         &   0.062 &   0.084 &         &         &       \\
2998124 & 28.598 &    NP & 56.785 & 0.717 &         &         &         &         &         &         &         &         &     a \\
3003991 &  7.245 &    SP &  9.563 & 0.493 &         &         &         &         &         &         &         &         &       \\
3097352 &  4.030 &    SP & 27.871 & 0.435 &   3.957 &   3.989 &         &         &   0.012 &   0.012 &   0.005 &   0.005 &     b \\
3098194 & 30.477 &    SP & 29.731 & 0.147 &  26.521 &  33.270 &         &         &   0.023 &   0.036 &         &         &       \\
3102000 & 57.060 &    SP & 14.733 & 0.905 &  13.998 &  15.483 &         &         &   0.048 &   0.043 &         &         &       \\
3102024 & 13.783 &    SP &  4.884 & 0.803 &         &         &         &         &         &         &         &         &       \\
3104113 &  0.847 &    EV &        &       &         &         &         &         &         &         &         &         &       \\
3113266 &  0.996 &    NP &  0.981 & 0.236 &         &         &         &         &         &         &         &         &     a \\
3114667 &  0.889 &    SP &  0.879 & 0.626 &         &         &         &         &         &         &         &         &       \\
3115480 &  3.694 &    SP &  3.617 & 1.198 &         &         &         &         &         &         &         &         &       \\
3119295 &  0.440 &    EV &        &       &         &         &         &         &         &         &         &         &       \\
3120320 & 10.266 &    SP & 13.261 & 0.782 &  12.473 &  13.670 &  14.389 &  14.617 &   0.028 &   0.032 &   0.031 &   0.043 &       \\
3122985 &  0.993 &    SP &  1.471 & 1.416 &   1.453 &   1.465 &   1.497 &   1.504 &   0.020 &   0.043 &   0.092 &   0.056 &       \\
3124420 &  0.949 &    EV &        &       &         &         &         &         &         &         &         &         &       \\
3127817 &  4.327 &    EV &        &       &         &         &         &         &         &         &         &         &       \\
3127873 &  0.672 &    EV &        &       &         &         &         &         &         &         &         &         &       \\
3128793 & 24.679 &    SP & 66.307 & 0.986 &         &         &         &         &         &         &         &         &       \\
3218683 &  0.772 &    EV &        &       &         &         &         &         &         &         &         &         &       \\
3221207 &  0.474 &    EV &        &       &         &         &         &         &         &         &         &         &       \\
\enddata
\tablecomments{A full version of the table is available in the online supplement.}
\tablenotetext{a}{$P_{ACF}$ and $h_{ACF}$ for the NP (no periodic out-of-eclipse variability) category are for validation purposes only, and should not be used for tidal synchronization analysis.}{b}{\change{$P_{ACF}$ is incorrect due to systematic artifacts in the light curve.}}
\label{tab:rotation catalog}
\end{deluxetable*}

Table \ref{tab:rotation catalog} lists a representative subset of entries in our rotation period catalog. The full catalog is available in the online supplement. For each EB, the table includes the orbital period and the visual classification. For EBs for which rotation periods were measured (category SP), the table lists the ACF rotation periods, ACF peak heights, as well as the periodogram periods and peak heights. ACF rotation periods and peak heights are listed for the nonperiodic out-of-eclipse variability (NP) category for validation purposes, but \change{flagged ``a'' in the Notes column} to indicate that they should not be used for tidal synchronization analysis. \change{For 12 EBs in the SP category flagged with ``b'', $P_{ACF}$ should not be used, however the periodogram periods are correct. The ACF detected a spurious signal due to systematic artifacts in the light curve.}

Unless otherwise stated, the following analysis uses the minimum periodogram-based rotation period for each EB (column $P_{1,min}$ in Table \ref{tab:rotation catalog}). \change{Assuming solar-like differential rotation, $P_{1,min}$ will be closest to the equatorial rotation period. This provides a consistent reference point for the differential rotation discussion below.}

\change{We also note that the conclusions presented below are the same when using the ACF-based periods, and so using the periodogram period does not bias our results. However, the periodogram based rotation periods provide more information than the ACF-based periods with regards to EBs with multiple rotation periods.}

\subsection{Orbital Period Distribution of EB Categories}

% Figure 5: P_orb histogram by class
\begin{figure}[ht]
\centering
\includegraphics[width=3in]{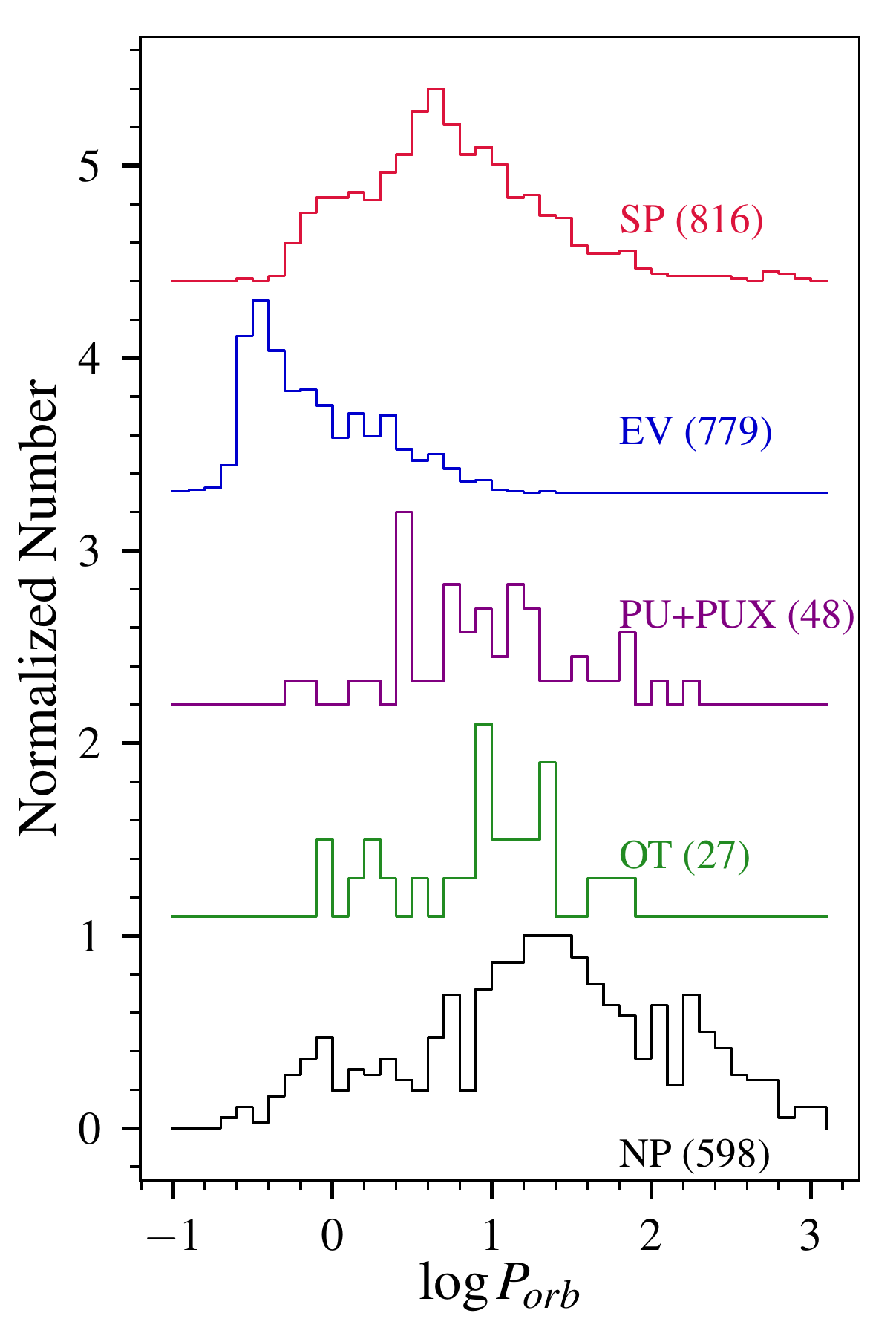}
\caption{The distributions of orbital periods for the light curve visual classifications. From top to bottom: starspot modulations (SP), ellipsoidal variations (EV), likely and possible pulsators (PU and PUX), other out-of-eclipse variability (OT), and no periodic out-of-eclipse variability (NP). \change{Each histogram has been normalized by its maximum value, and the histograms are vertically offset for clarity.} The number of EBs in each class is indicated in parentheses.}
\label{fig:class p_orb hist}
\end{figure}

Figure \ref{fig:class p_orb hist} shows the distributions of orbital periods for the five true EB categories from $\S$\ref{subsec:classification}, not including the last category (10 objects) where starspots may have been mistaken for ellipsoidal variations. \change{The distributions show} evidence of tidal interaction. Strong tidal forces at short orbital periods drive the ellipsoidal variations. Compared to the non-periodic category, EBs with starspot modulations favor shorter orbital periods where the stars are tidally spun up, resulting in stronger magnetic activity. The non-periodic systems are concentrated at longer orbital periods where the tidal forces are weaker. These EBs have not synchronized, so the stars are rotating more slowly and therefore do not have strong magnetic activity that produces detectable starspot modulations. The pulsation and other variability categories do not show a strong dependence on orbital period, because these processes are apparently independent of rotation and hence orbital period.

\subsection{The Period Ratio Diagram}
\label{subsec:p_orb-p_rot diagram}

To measure the degree of synchronization for a given EB, we compute the \textit{period ratio} $P_{orb}/P_{rot}$. This is equal to $\Omega_{\star}/n$, where $\Omega_{\star}$ is the rotational angular velocity of the star, and $n$ is the mean orbital angular velocity. Synchronization occurs at $P_{orb}/P_{rot} = 1$, while  $P_{orb}/P_{rot} > 1$ is supersynchronous, $P_{orb}/P_{rot} < 1$ is subsynchronous.

\begin{figure}[ht!]
\centering
\includegraphics[width=3.4in, trim = 0cm 0cm 0cm 0 cm, clip]{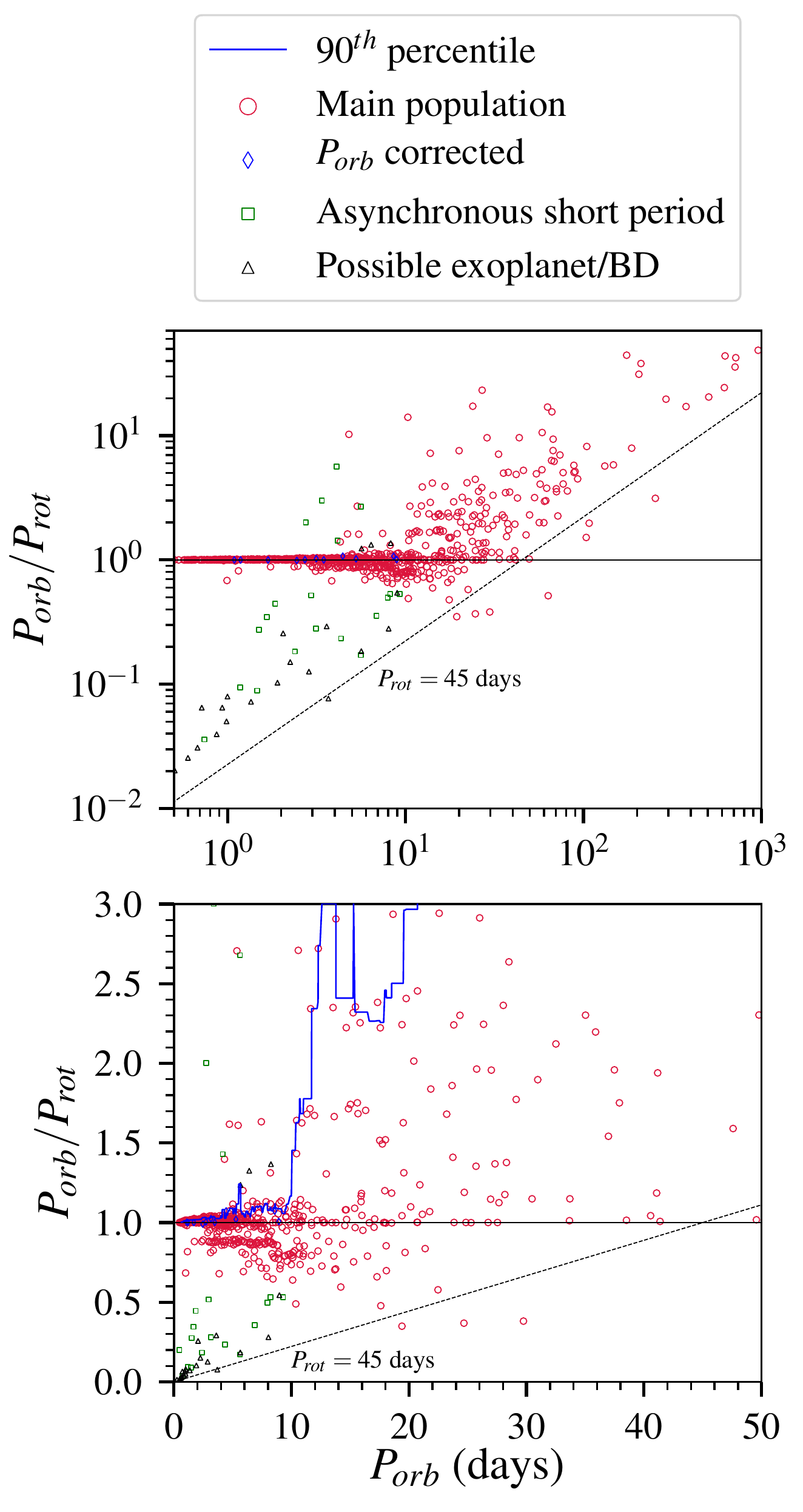}
\caption{The distribution of period ratio versus orbital period for the EBs with starspot modulations. Likely non-EB outliers are indicated by black triangles, EBs with orbital period corrections are indicated by blue diamonds, and asynchronous short period EBs are indicated by green squares. The \change{black horizontal} line corresponds to synchronization at $P_{rot} =  P_{orb}$, while the dashed \change{diagonal} line indicates conservative rotation period limit of 45 days. \change{The blue curve indicates the running \th{90} percentile.} The bottom panel shows the region around synchronization in more detail.}
\label{fig:p_rot vs p_orb sp}
\end{figure}

Figure \ref{fig:p_rot vs p_orb sp} shows the \change{period ratio diagram} for the 816 EBs in the SP category. \change{These EBs are divided into a main population, and three categories of outliers. The outliers are discussed below, before moving on to the main population.}

% Figure 6: P_orb / P_rot vs. P_orb for SP

\subsubsection{Asynchronous Systems with Short Periods}

\label{subsubsec:outliers}

Before investigating trends in synchronization, we identify 61 asynchronous systems with orbital periods less than 10 days that have a period ratio less than 0.6 or greater than 1.2. The outliers are listed in Table \ref{tab:outliers} in the Appendix, and are divided into four categories.

\begin{enumerate}

  \item There are 11 EBs where the rotation period is exactly twice or half of the orbital period. We argue that these are not in fact outliers, but instead the KEBC orbital period is incorrect. This can occur because it is difficult to distinguish between a circular EB with only primary eclipses, and an EB with nearly equal primary and secondary eclipse depths at twice the period. Our rotation period measurement could be incorrect by a factor of two due to aliasing effects, but this is unlikely as we used the ACF for validation. We therefore corrected the orbital periods, moving these EBs into the synchronized population. They are indicated by blue diamonds in Figure \ref{fig:p_rot vs p_orb sp}.
  
  \item There are 21 systems with unambiguous primary and secondary eclipses, meaning they are most likely EBs. These EBs may be asynchronous because they are young, or have a complex dynamical history. They are indicated by green squares in Figure \ref{fig:p_rot vs p_orb sp} and are included in the synchronization analysis below.
    
  \item There are 22 systems that are likely not EBs. They are indicated by black triangles in Figure \ref{fig:p_rot vs p_orb sp}, and are \textit{not} included in the analysis below. We further divide these systems into two categories:

  \begin{enumerate} 
    \item There are 12 systems with very low signal-to-noise primary eclipses and no secondary eclipses. They may be false positives because a close, stellar mass companion should have synchronized the binary.
    
    \item There are 10 systems with unambiguous but shallow primary eclipses and no secondary eclipses. The occulting object may be a planet or brown dwarf, which is not massive enough to have synchronized the star. \newline
    
    \indent Of these 22 systems, \citet{Kolbl2015} found that KIC 7763269, KIC 9752973, KIC 10338279, and KIC 10857519 show evidence of a close stellar companion in their spectra. However, without multi-epoch radial velocities, it is unclear whether the spectral companion is responsible for the eclipses.

  \end{enumerate}
  
  \item There are 7 EBs in this range ($P_{orb} < 10$ days and $0.6 < P_{orb}/P_{rot} < 1.2$) that appear to be pseudosynchronized, as discussed in $\S$\ref{subsubsec:pseudosync}.
  
\end{enumerate}

Furthermore, it is possible that the starspot modulation we detected does not originate from the EB at all, and instead comes from a third star in the system, or an unrelated star at a small angular separation. All of these outlying systems are worthwhile targets for observational followup, especially those that are potentially young or have an interesting dynamical history. With a small number of radial velocity and/or adaptive optics observations, it would be straightforward to distinguish between the cases listed above.

\subsection{Dependence on Orbital Period}
Orbital period is arguably the most important quantity for tidal synchronization, as the synchronization timescale is predicted to increase with orbital period to the the sixth power \citep{Hut1981}.

\label{subsec:orbital period}

\subsubsection{Synchronization and Differential Rotation Below 10 Days}

As seen in Figure \ref{fig:p_rot vs p_orb sp}, EBs with orbital periods less than 2 days are nearly all synchronized. 94\% of the sample has $0.92 < P_{orb}/P_{rot} < 1.2$. Between 2 and 10 days, the sample is divided into two clusters. The main cluster is centered slightly above the synchronization line, while the second cluster is centered around $P_{orb}/P_{rot} = 0.87$. 72\% of EBs with orbital periods between 2 and 10 days have have $0.92 < P_{orb}/P_{rot} < 1.2$ (main cluster), while 15\% have $0.84 < P_{orb}/P_{rot} < 0.92$ (subsynchronous cluster).

\change{The subsynchronous rotation periods of the EBs is not an instrumental or numerical artifact. The subsynchronous peaks are present in the full light curve periodogram (black curve in lower left panel of Figure \ref{fig:eclipse removal}), meaning that interpolating over the eclipses cannot explain the subsynchronous rotation. Furthermore, the ACF-based rotation periods also have a subsynchronous cluster, so this cannot be an artifact of the periodogram. We therefore conclude that the subsynchronous signal is due to starspot modulations.}

The \change{cluster} of subsynchronous EBs are an unexpected and intriguing result. To our knowledge, this phenomenon has not been observed previously. In $\S$\ref{sec:diff rot}, we demonstrate that the subsynchronous rotation is consistent with differential rotation. If the stars are tidally synchronized at the equator, then starspots at higher, slower rotating latitudes will make the measured rotation period subsynchronous. 

\subsubsection{A Transition to Eccentric, Pseudosynchronized EBs}
\label{subsubsec:pseudosync}

Beyond roughly 10 days there is a decrease in the number of EBs centered around the synchronization line. This coincides with an increase in the number of supersynchronous EBs ($P_{orb}/P_{rot} > 1.2$).

To quantify this transition, we compute the \th{90} percentile of the period ratio distribution in a running manner.  For each EB, we take the other 29 EBs with the nearest orbital periods to calculate the percentile. \change{The asynchronous non-EB systems ($\S$\ref{subsubsec:outliers}) were excluded.} A larger value of the \th{90} percentile indicates that the supersynchronous tail of the distribution is more significant.

\change{The running \th{90} percentile is plotted as a thick black curve in the bottom panel of Figure \ref{fig:p_rot vs p_orb sp}.} At 10 days there is a rapid increase in the \th{90} percentile. Although there are some supersynchronous EBs at shorter periods, they are a small fraction of the sample, whereas at 10 days the fraction of supersynchronous EBs increases dramatically at the expense of synchronized EBs.

We therefore divide the period ratio distribution into two main populations. Below 10 days, the bulk of EBs are synchronized, with a subpopulation of subsynchronous rotators. Above 10 days, there is a significant increase in the number of supersynchronous rotators. As we demonstrate below, this transition occurs because a large fraction of the EB orbits are eccentric, and those EBs are pseudosynchronized. 

\begin{figure}[ht!]
\centering
\includegraphics[width=3.4in, trim = 0cm 0cm 0cm 0cm, clip]{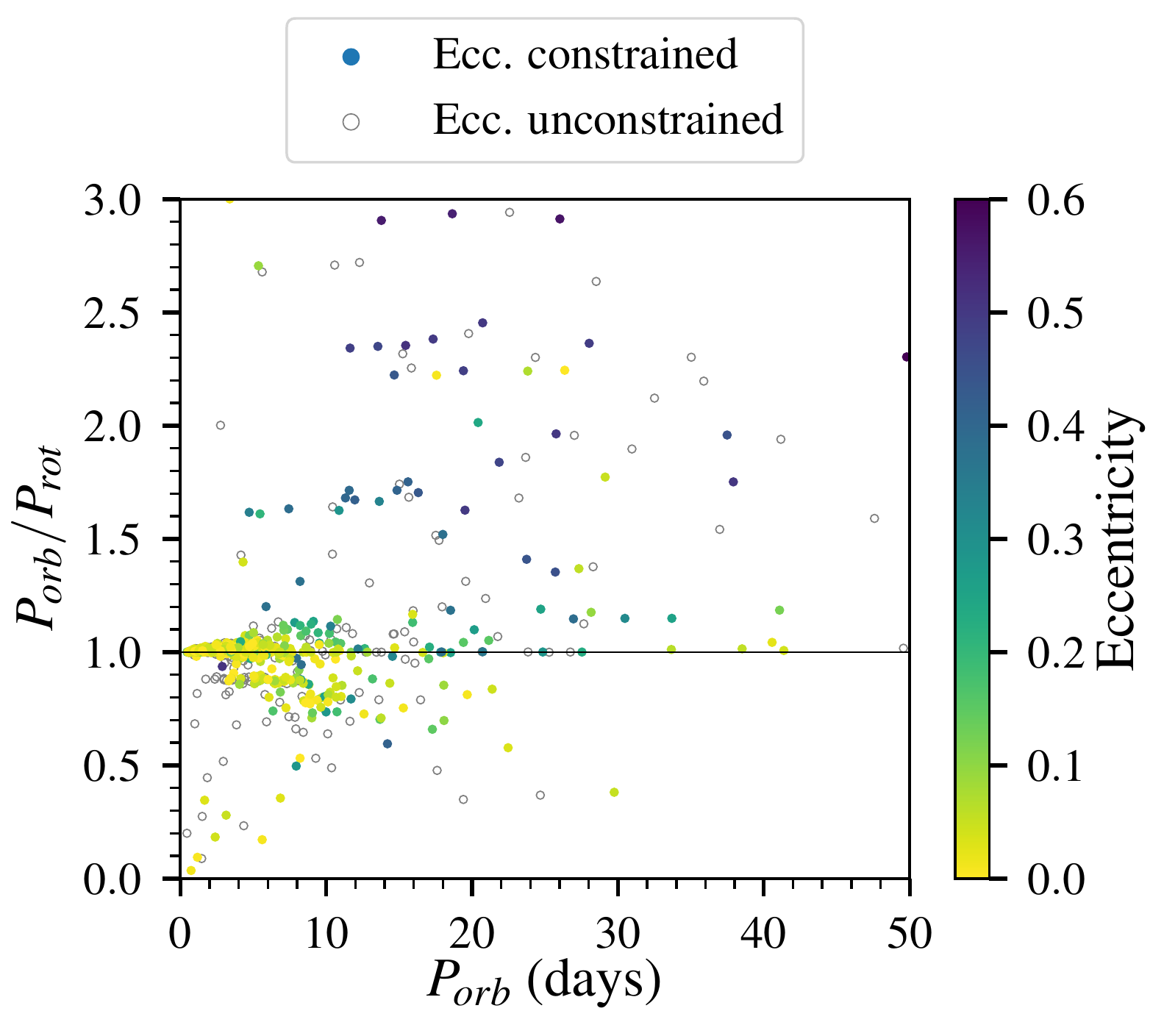}
\caption{Distribution of period ratio versus orbital period for EBs with starspot modulations. Points are colored according to eccentricity. EBs without eccentricity constraints are indicated by open grey circles.}
\label{fig:p_orb p_rot eccentricity}
\end{figure}

Figure \ref{fig:p_orb p_rot eccentricity} shows the same distribution as \change{Figure \ref{fig:p_rot vs p_orb sp}}, but the points are colored according to eccentricity, measured as described in $\S$\ref{subsec:eccentricities}. There is a clear division in Figure \ref{fig:p_orb p_rot eccentricity} based on eccentricity. Most of the EBs with small eccentricities (yellow circles) have orbital periods less than 10 days and are concentrated near synchronization. In contrast, most of the EBs with larger eccentricities (dark green and purple circles) have orbital periods greater than 10 days, and are supersynchronous.

% Figure 7: P_orb/P_rot with eccentricity

The distribution of EBs without eccentricity constraints overlaps those with constraints. If the eccentricities were measured, it is reasonable to assume that they would follow the same trends described above. Alternatively, some supersynchronous EBs without eccentricity constraints may not have tidally interacted. The EB may have a low mass ratio, as is consistent with a lack of secondary eclipses, which are required to measure eccentricity. 

Binaries in eccentric orbits are expected to become ``pseudosynchronized'', so that the rotational angular velocity is nearly equal to the instantaneous orbital angular velocity at periastron. A pseudosynchronized EB would appear supersynchronous in our sample because the orbital angular velocity at periastron is greater than the mean orbital angular velocity.

Pseudosynchronization can explain the slightly supersynchronous rotation of the main cluster of EBs with periods less than ten days. These slightly supersynchronous EBs may have eccentricities that are too small to measure by our approximation. In that case they would technically be pseudosynchronized, but only slightly supersynchronous due to the small eccentricity. Consistent with this scenario, the upper right corner of the cluster has the largest eccentricities (light green points), and are also the most supersynchronous. This is unlikely to be an artifact of the periodogram analysis, because the ACF-based rotation periods are also slightly supersynchronous.

\begin{figure}[ht!]
\centering
\includegraphics[width=3.4in, trim = 0cm 0cm 0cm 0cm, clip]{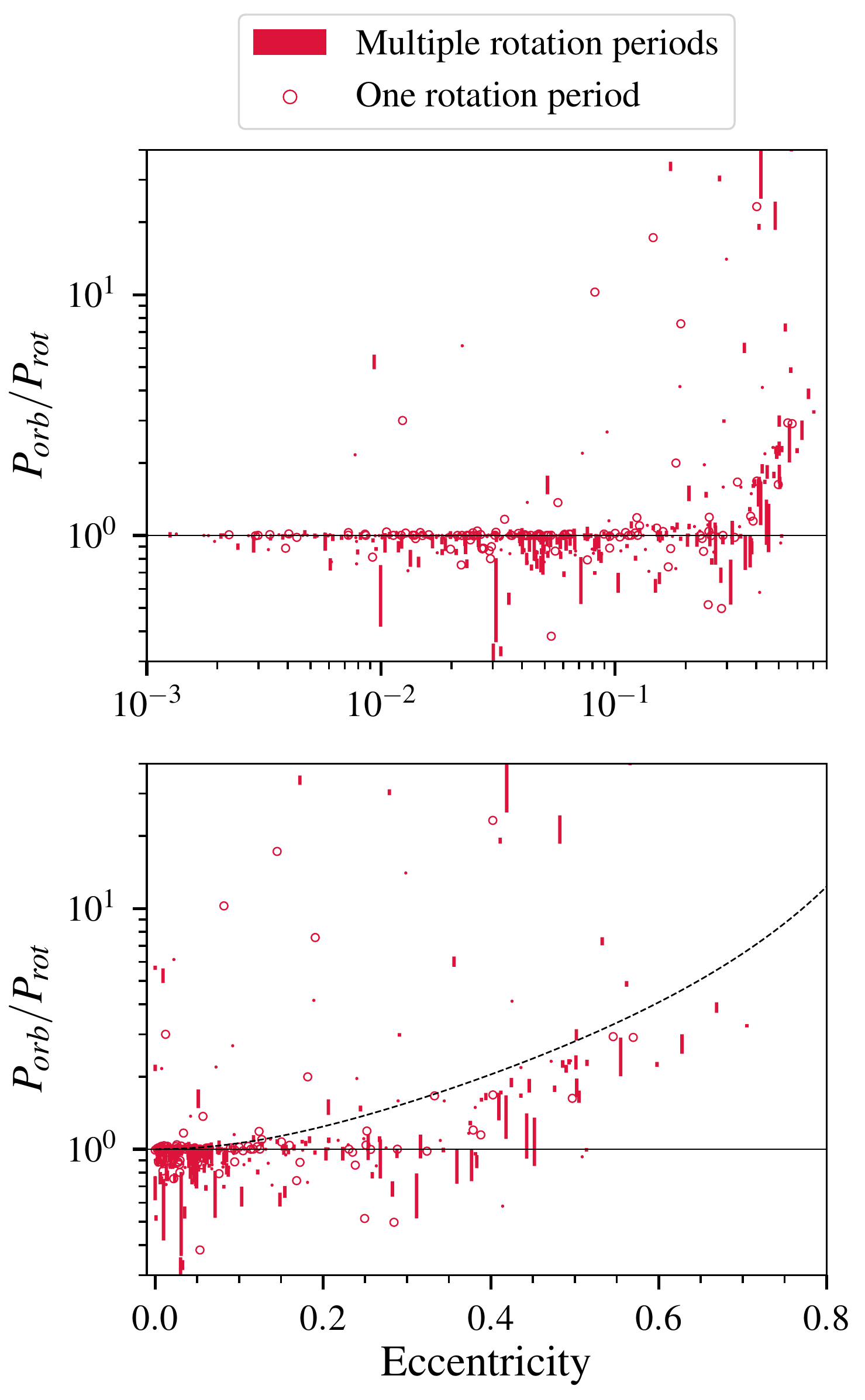}
\caption{The distribution of period ratio versus eccentricity for EBs with starspot modulations. Vertical bars indicate the range of orbital periods measured, while open circles indicate EBs with only one rotation period measurement. The solid line corresponds to synchronization, while the dashed curve shows the predicted value of the period ratio from \citet{Hut1981} for pseudosynchronization.}
\label{fig:p_rot_ecc}
\end{figure}

Further evidence for pseudosynchronization is found in the distribution of the period ratio versus eccentricity, shown in Figure \ref{fig:p_rot_ecc}. The eccentric EBs appear to be pseudosynchronized, but are below the model prediction \change{of \citet[][Eq. 42]{Hut1981}} by up to 50\%. \change{Of the four EBs in our sample with eccentricity measurements by \citet{Kjurkchieva2016} and \citet{Kjurkchieva2018}, three agree to within 5\%, and one we overestimate by 26\%.} In $\S$\ref{sec:diff rot}, we argue that this may be due to differential rotation. Alternatively, the model may underpredict the pseudosynchronization period

% Figure 8: P_orb/P_rot vs ecc.

\citet{Zahn1989} predicted the existence of a cutoff orbital period for circularization between 7.2 and 8.5 days for stars with masses between 0.5 and 1.5 $M_{\odot}$. The cutoff is determined by the maximum orbital period at which the extended pre-main sequence binaries can circularize. When the stars begin to contract onto the main sequence, the rotation rate increases and becomes supersynchronous, but the orbit remains circular. Binaries that do not circularize on the pre-main sequence slowly circularize during the main sequence phase. \citet{Meibom2005} report a tidal circularization period of $10.3^{+1.5}_{-3.1}$ days based on data for 50 nearby solar-type binaries from \citet{Duquennoy1991}. This is in excellent agreement with the rapid increase in eccentric, supersynchronous EBs near 10 days.

Previous studies support the existence of a transition period for pseudosynchronization. \citet{Mazeh2008} compiled data for eight pre-main sequence binaries from \citet{Marilli2007} and six binaries in young clusters from \citet{Meibom2006}. Orbital parameters were determined from radial velocities, and rotation periods were determined from starspot modulations. \citeauthor{Mazeh2008} finds a transition period between 8 and 10 days from circular, synchronous binaries to eccentric, supersynchronous binaries, in excellent agreement with our result.

Of the seven binaries that are eccentric and supersynchronous in \citeauthor{Mazeh2008}'s sample, the two most eccentric binaries are rotating slower than the predicted pseudosynchronization period, while the other five are rotating faster than predicted. This is in contrast to our sample, where the majority of eccentric binaries are rotating slower than predicted, with the caveat of differential rotation discussed previously. \citeauthor{Mazeh2008} argues that the stars in the compiled sample are too young to have achieved pseudosynchronization, and that an older population of binaries would show a greater degree of pseudosynchronization. The latter appears to have occurred for our sample of Milky Way field binaries.

It is possible that some EBs are in a spin-orbit resonance. Unlike planets such as Mercury, stars do not have a fixed shape that would lead to a resonance. However, the existence of eccentric, supersynchronous EBs leaves open the possibility for coupling with the convective motions or internal pressure and gravity modes \citep{Burkart2014}. \change{There is no obvious clustering of EBs near the 2:1 or 3:2 resonances ($P_{orb}/P_{rot} = 2$ and 1.5), however there is some suggestion of clusters near $P_{orb}/P_{rot}$ = 1.6 and 2.3. The nearest, low integer ratio resonances are 5:3 and 7:3}, although we hesitate to draw any conclusions given the small number of EBs in this range. 

\subsubsection{Behavior at Longer Periods}
\label{subsubsec:longer periods}

So far, we have focused on eccentric, pseudosynchronized binaries. However, Figure \ref{fig:p_orb p_rot eccentricity} also contains some EBs with small eccentricities and orbital periods greater than 10 days that are synchronized or nearly synchronized. This raises the question of to what extent circularization and synchronization continue during the main sequence phase.  The EBs in our sample are part of the Milky Way field population. They should typically be at least a few Gyr old, and therefore have had a long main sequence phase during which tidal interaction could take place.

While old binaries are circularized at longer periods than young binaries, the difference is only about a factor of two. \citet{Latham2002} reported orbital solutions for 171 high proper motion binaries, which are likely members of the halo. For this sample, \citet{Meibom2005} found a circularization period of $15.6^{+2.3}_{-3.2}$ days. Thus even for the oldest main sequence binaries in the Galaxy, we should not expect tidal circularization to have reached beyond $\sim$20 days.

Our results support this conclusion, in that we observe very few synchronized EBs with small eccentricities and orbital periods longer than 10 days. Five notable exceptions seen in Figure \ref{fig:p_orb p_rot eccentricity} are synchronized and have nearly circular orbits between 32 and 50 days. These are KIC 3955867, KIC 4569590, KIC 5308778, KIC 7133286, and KIC 8435232. They have flat-bottomed primary and secondary eclipses, which we interpret as containing a main sequence and evolved star. If one of the stars is evolved, its larger radius would allow for tidal circularization and synchronization at longer orbital periods than on the main sequence.

Beyond 30 days, there are very few synchronized EBs (except the possibly evolved stars), and only a handful of possibly pseudosynchronized EBs. This is consistent with the expectation that tidal interaction decreases rapidly with increasing orbital period.

\subsection{Dependence on Stellar Mass (Color)}

We now investigate the dependence of tidal synchronization on stellar mass. For a given semimajor axis, the synchronization timescale decreases with stellar radius to the sixth power \citep{Hut1981}. We therefore expect that EBs with more massive primaries (larger radii) should be synchronized at longer periods. However, the timescale also depends on other factors, including the mass ratio (see $\S$\ref{subsec:mass ratio}) and initial eccentricity. Furthermore, the efficiency of the tidal dissipation mechanism likely depends on the thickness of the convective envelope, which increases with decreasing mass.

\change{Photometric colors are the only mass estimates available for the entire sample. In what follows, we assume that the EBs contain main sequence stars (with the exception of the five possibly evolved stars noted above), and that $g-K$ colors from the \Kepler Input Catalog \citep{Brown2011} are indicative of the mass of the primary star. As a conceptual tool, Table \ref{tab:spectral types} divides the sample into spectral types A through M, using the main sequence color relations from \citet{Covey2007}. Prior to assigning spectral types, we corrected for interstellar reddening using the $E(B-V)$ values in the Kepler Input Catalog. We stress that these spectral types are intended as approximations, given the limited mass information available for most of the sample.}

\begin{figure}[ht]
\centering
\includegraphics[width=3.4in, trim = 0cm 0cm 0cm 0cm, clip]{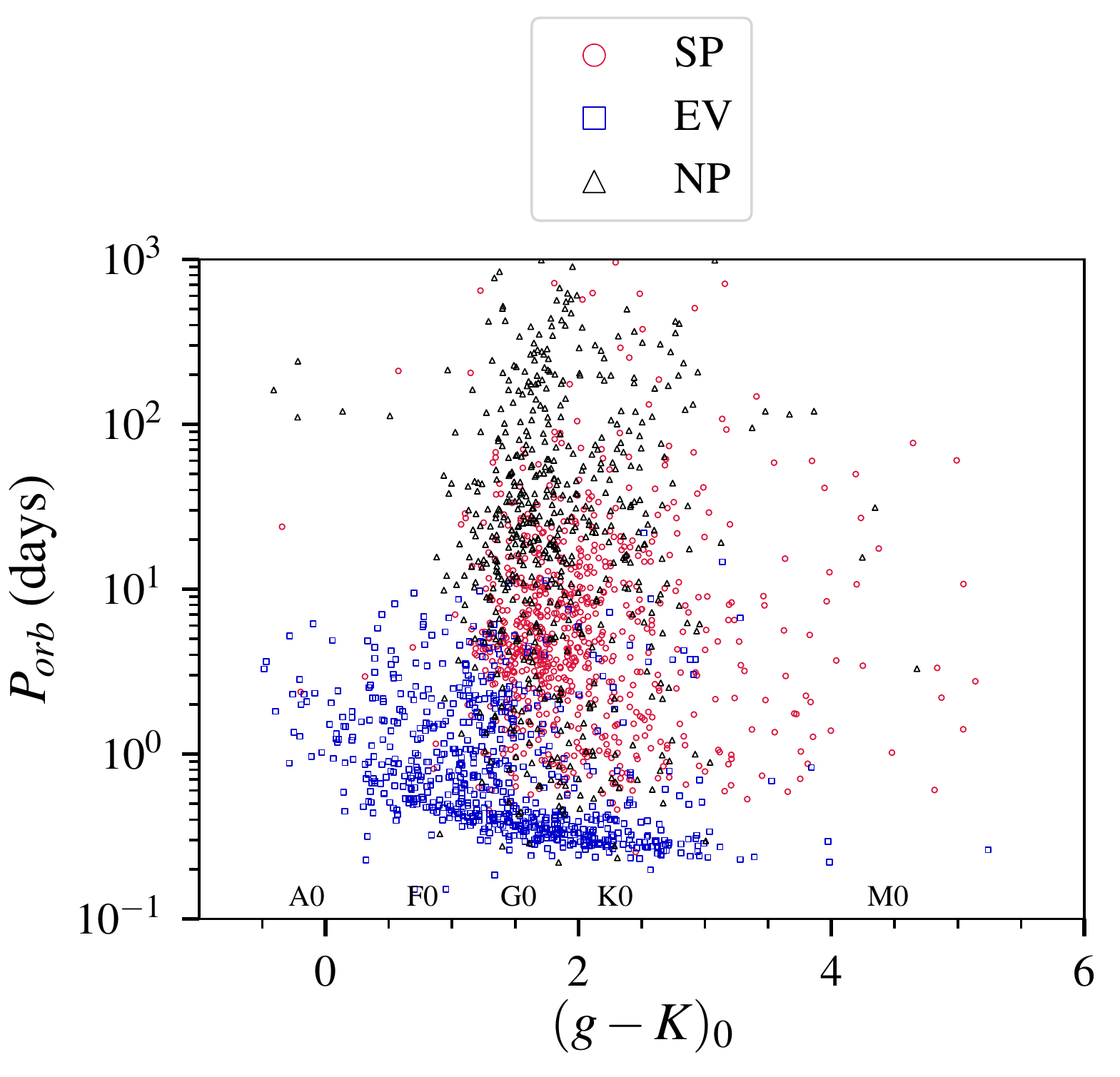}
\caption{The distribution of orbital period versus \change{dereddened} $g-K$ color for EBs with starspot modulations (red circles), ellipsoidal variations (blue squares), and non-periodic out-of-eclipse variability (black triangles). Spectral types from \citet{Covey2007} are given for reference.}
\label{fig:p_orb_teff}
\end{figure}

Figure \ref{fig:p_orb_teff} shows the distribution of orbital periods versus \change{dereddened} $g-K$ color. \change{Over half (57\%)} of the EBs with A and F primaries have ellipsoidal variations. A and F stars are not expected to have starspots, which would leave ellipsoidal variations as the dominant source of out-of-eclipse variability. For the cluster of short period ellipsoidal variables, there is a trend of decreasing $P_{orb}$ with \change{increasing $g-K$ values}, which we discuss in $\S$\ref{subsec:contact variables}.

% Figure 9: P_orb vs g-K

% Table 3: Spectral Types of Rotation Period Catalog
\begin{deluxetable}{llll}

\tabletypesize{\small}
\tablewidth{0pt}
\tablecaption{Spectral Types of Rotation Period Catalog}

\tablehead{
\colhead{Sp. Type}&
\colhead{$g-K$}&
\colhead{Number} &
\colhead{Structure}}

\startdata
A & $< 0.8$ & \change{5} &  Radiative  envelope\\
F & 0.8 - 1.5 & \change{122} &  Small convective envelope \\
G & 1.5 - 2.3 & \change{428} &  Medium convective envelope \\
K & 2.3 - 4.5 & \change{181} &  Medium convective envelope \\
M0 - M4 & 4.5 - 6.2 & \change{8} &  Large convective envelope \\
\enddata

\tablecomments{$g-K$ colors are taken from \citet{Covey2007} for dwarfs. 8 EBs do not have $g-K$ values\change{, and 64 do not have $E(B-V)$ values listed in the Kepler Input Catalog, and so were not assigned spectral types.}}
% \tablenotetext{}{}
\label{tab:spectral types}
\end{deluxetable}

Nearly all EBs \change{(90\%)} have \change{F,} G and K primaries, reflecting the selection of solar-like stars for the \Kepler target list. This selection effect is beneficial in the sense that it greatly increases the number rotation period measurements for primaries with convective envelopes, whereas most previous observational studies of synchronization focused on primaries with radiative envelope.

Using the above color limits, the rotation period catalog only contains no fully convective primaries \change{(later than M4)}, and \change{five} primaries with radiative envelopes. These numbers are insufficient to draw any conclusions about the tidal synchronization changes in the radiative envelope and fully convective regimes. We therefore concentrate on the differences between \change{F,} G and K primaries.

\begin{figure}[ht]
\centering
\includegraphics[width=3in, trim = 0cm 0cm 0cm 0cm, clip]{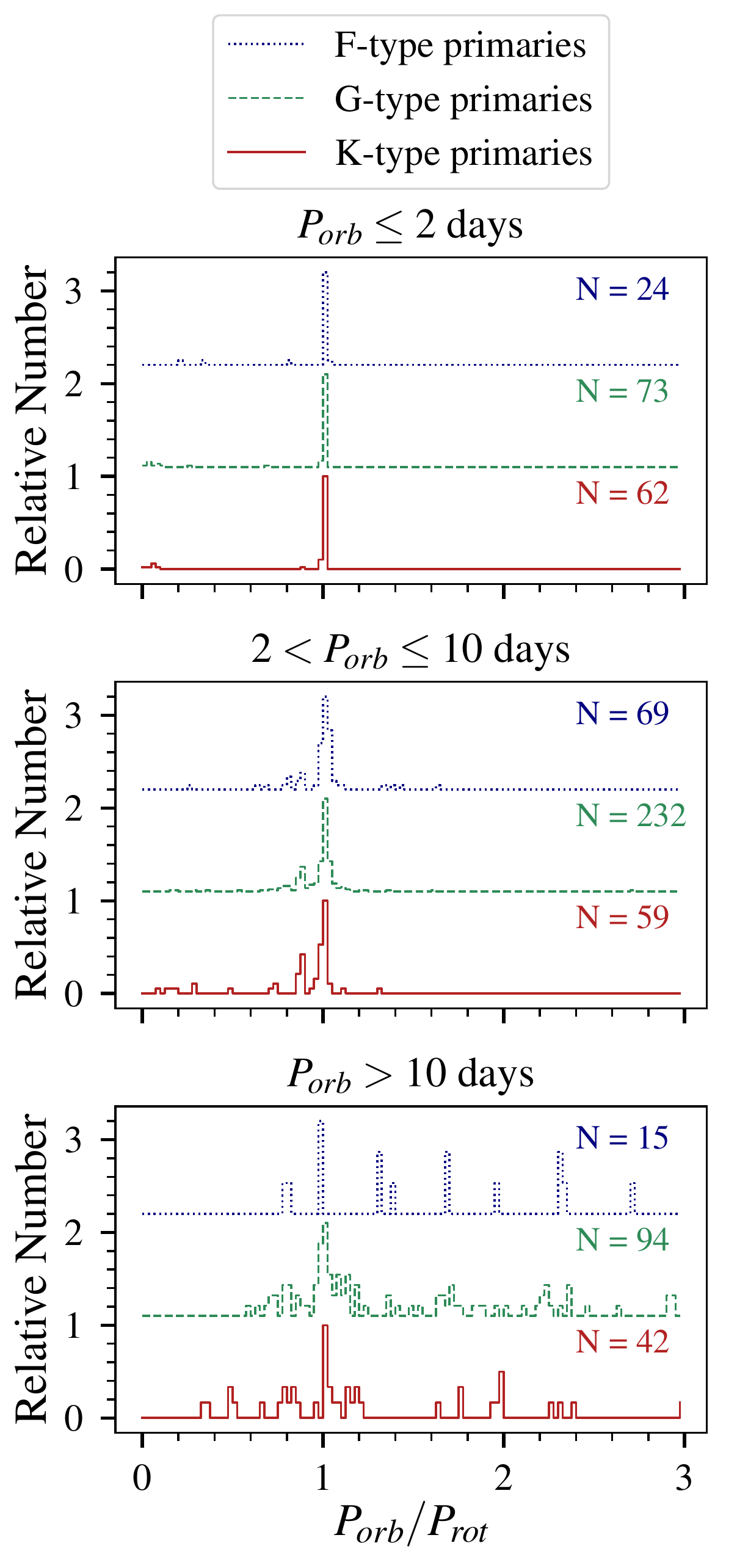}
\caption{From top panel to bottom: the distributions of period ratio for $P_{orb} \le 2$ days (top panel), $2 < P_{orb} \le 10$ days (middle), and $P_{orb} > 10$ days (bottom). EBs with \change{F-}, G-, and K-type primaries are denoted by \change{dotted blue}, dashed green and solid red lines, respectively. Each histogram is normalized to its maximum value and vertically offset for clarity. The number of EBs in each histogram is listed.}
\label{fig:sync_vs_teff}
\end{figure}

Figure \ref{fig:sync_vs_teff} shows the distributions of period ratio for three different orbital period ranges: $P_{orb} \le 2$ days, $2 < P_{orb} \le 10$, $P_{orb} > 10$, \change{with separate histograms for F,} G, and K primaries. Our results indicate that there is no obvious difference in the period ratio distribution over the relatively narrow mass and radius range spanned by \change{F,} G, and K primaries. Thus primary mass \change{does not appear to be} a strong factor in the tidal synchronization of \change{the F,} G, and K primaries \change{in our sample}.

% Figure 10: P_orb/P_rot histograms for G and K dwarfs
 
\subsection{Dependence on Mass Ratio}
\label{subsec:mass ratio}
Given the above results for primary mass, we now investigate the dependence of tidal synchronization on mass ratio, defined as $M_{sec}/M_{pri}$. The mass ratio has a maximum value of one for equal mass binaries, and approaches zero for very unequal masses. The tidal synchronization timescale is predicted to decrease with increasing mass ratio \citep{Hut1981}, so that EBs with nearly equal mass ratios should be synchronized at longer periods than EBs with low mass ratios, keeping all other factors constant.

We create two subsamples of EBs with the greatest difference in mass ratio. The first subsample has primary eclipse depths $\delta_{pri} < 0.1$, and no detected secondary eclipses, indicating a small companion mass relative to the primary. The second subsample has ratios of primary to secondary eclipse depth $\delta_{sec}/\delta_{pri} > 0.7$, indicating a roughly equal mass companion.

\begin{figure}[ht]
\centering
\includegraphics[width=3in, trim = 0cm 0cm 0cm 0cm, clip]{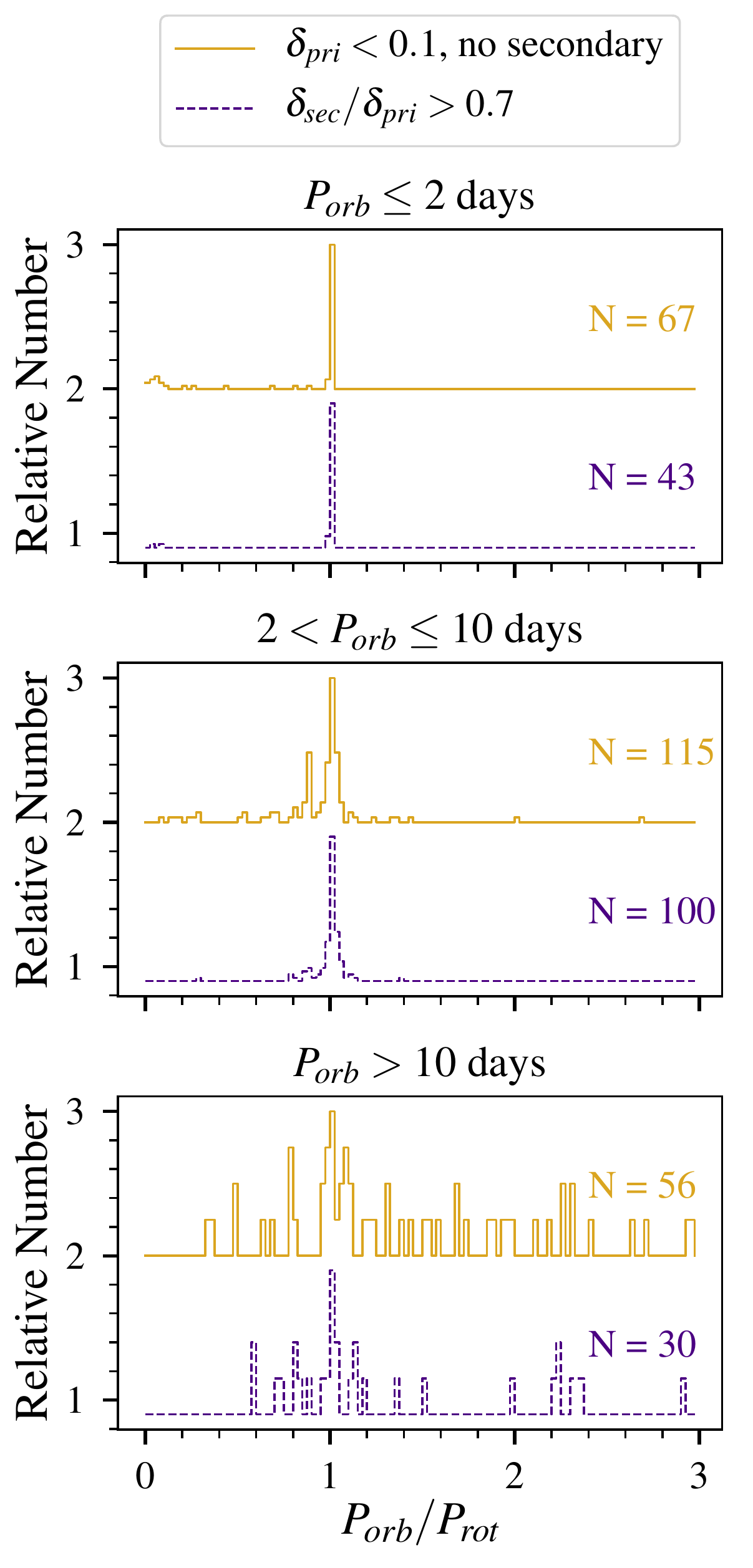}
\caption{The dependence of synchronization on the mass ratio. The distribution of the period ratio is shown for three orbital period ranges: $P_{orb} \le 2$ days (top panel), $2 < P_{orb} \le 10$ days (middle), and $P_{orb} > 10$ days (bottom). The solid yellow histograms are for EBs with primary eclipse depths less than 0.1, and no secondary eclipses. This indicates a small mass ratio. The dashed purple histograms are for EBs with secondary-to-primary eclipse depths ratios greater than 0.7, indicating a \change{roughly equal} mass ratio.}
\label{fig:sync_vs_depth}
\end{figure}

Figure \ref{fig:sync_vs_depth} shows the distributions of period ratio for $P_{orb} \le 2$ days, $2 < P_{orb} \le 10$ days, and $P_{orb} > 10$ days \change{, with separate histograms for the small and roughly equal mass ratio subsamples.} Most EBs with orbital periods less than 2 days are synchronized. This is true regardless of the mass ratio, although there are some subsynchronous EBs with low mass ratios as discussed in $\S$\ref{subsubsec:outliers} (small peak near zero in top yellow histogram). In the 2 to 10 day orbital period range, the low mass ratio EBs have a higher relative number of $\sim$13\% subsynchronous EBs compared to the equal mass ratio subsample. At orbital periods longer than 10 days, the equal mass ratio subsample is somewhat more synchronized than the low mass ratio subsample, with 44\% of the equal mass ratio subsample having rotation periods within 20\% of the orbital period, compared to 22\% for low mass ratio EBs.

% Figure 11: P_orb/P_rot distributions vs mass ratio

It appears that synchronization has a somewhat stronger dependence on mass ratio than on the mass of the primary. However, the mass ratio of our sample spans a relatively narrow range from 1 to roughly 0.1, because the companions are likely stars. Some systems may have substellar companions and be asynchronous ($\S$\ref{subsubsec:outliers}), suggesting that mass ratio becomes more important in the very small mass ratio regime.

\section{Differential Rotation}
\label{sec:diff rot}

As was noted repeatedly in the previous section, there is a population of subsynchronous EBs with orbital periods between two and ten days. Additionally, there is a population of eccentric EBs that are rotating supersynchronously, as is consistent with pseudosynchronization, but that are rotating up to 50\% slower than predicted by the model of \citet{Hut1981}.

In this section, we argue that both of these populations can be explained by differential rotation. We first examine the differential rotation measurements of the EBs, and conclude that they are consistent with single stars. Then we demonstrate how differential rotation explains the observed subsynchronous rotation.

\subsection{Comparison to Single Stars}
\label{subsec:multiple period results}

Of the 816 stars with starspot modulations, 206 had two periodogram peak groups, while 422 had one peak group. The remaining 188 only had a single significant peak, and hence do not show evidence of multiple rotation periods.

Following RG15, we express differential rotation in two ways. Absolute shear $d\Omega = 2 \pi (1/P_{min} - 1/P_{max})$ measures the difference in rotational frequency between two latitudes in radians per day. $P_{max}$ and $P_{min}$ are the maximum and minimum rotation periods identified in $\S$\ref{subsubsec:multiple periods}. On a star with $d\Omega = 0.05$ rad/day, the slower rotating latitude would lag the faster latitude by 0.05 rad $=2.86^{\circ}$ after one day. This quantity is measured directly from the frequency difference in the periodogram peaks. 

Relative shear is defined as $\alpha = (P_{max} - P_{min}) / P_{max}$. This is equal to the difference in rotation period between the poles and equator relative to the poles, and can take values between zero and one. Relative shear is a more intuitive quantity to understand the subsynchronous rotation scenario in $\S$\ref{subsec:subsynchronous}. Our differential rotation measurements are lower limits, because the starspots that trace rotation may not be exactly on the equators and poles. 

% Figure 12: Relative shear versus rotation period
\begin{figure}[ht]
\centering
\includegraphics[width=3in, trim = 0cm 0cm 0cm 0cm, clip]{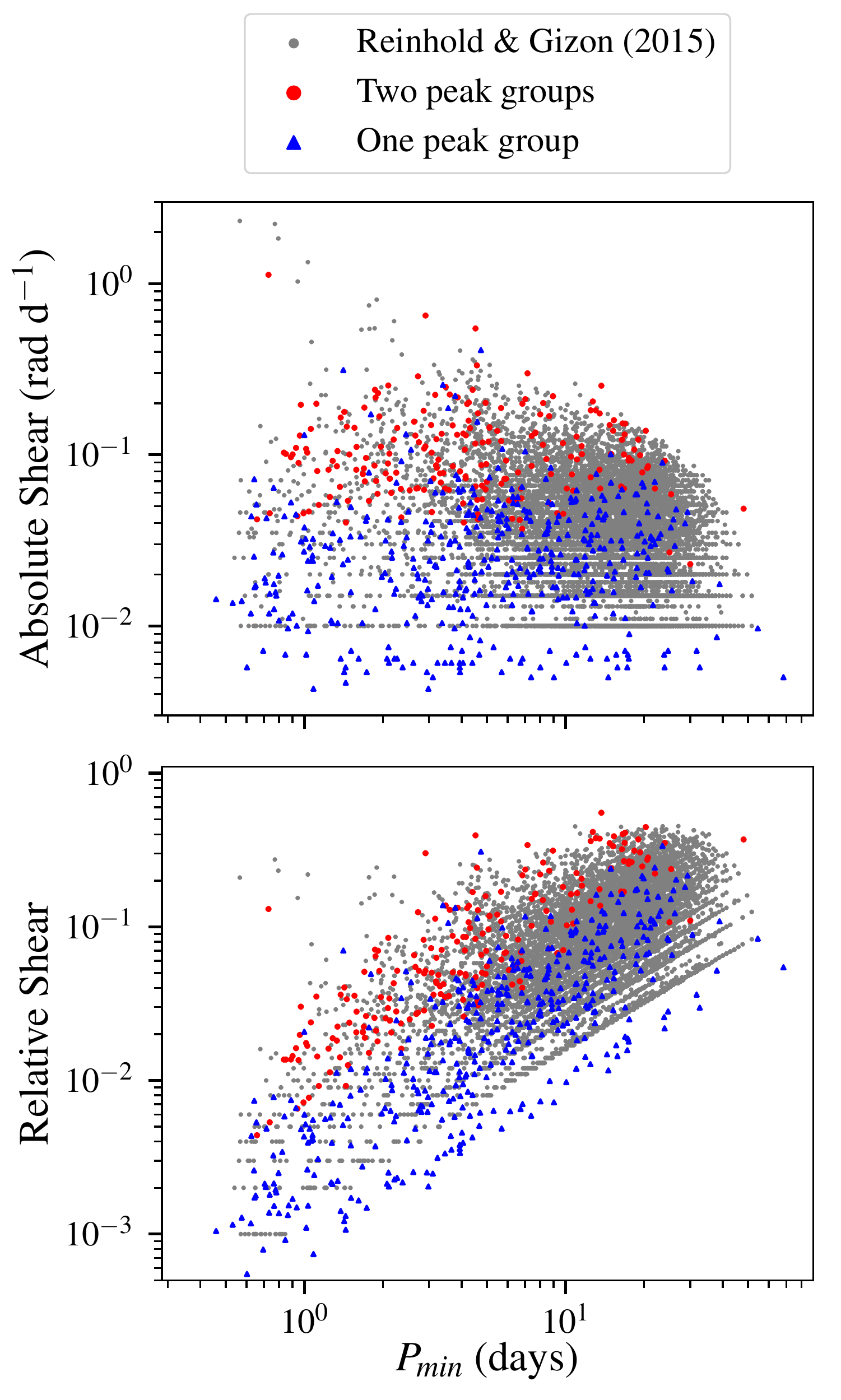}
\caption{Absolute shear $d\Omega$ (top panel) and relative shear $\alpha$ (bottom panel) versus the minimum periodogram rotation period. EBs with two groups of peaks are shown as red circles, and EBs with one peak group as blue triangles. For comparison, the single star sample of \citet{Reinhold2015} for $T_{eff} < 6300$ K is shown as grey circles.}
\label{fig:shear vs p_rot}
\end{figure}

The top and bottom panels of Figure \ref{fig:shear vs p_rot} show the distribution of absolute and relative shear versus the minimum rotation period measured for each EB. For comparison, \change{we show} the single star sample of RG15 with $T_{eff} < 6300$ K.

In general, our sample overlaps with the RG15 sample. The sequence of blue triangles below the RG15 distribution is most likely due to differences between our periodogram analyses. We therefore conclude that the vast majority of the multiple rotation period results are consistent with differential rotation of starspots detected on the only the primary star.  Notable exceptions are KIC 10068919, KIC 11147460, and KIC 11231334, which have shear measurements above the RG15 sample, and are the best candidates for having periods originating from the two separate stars in the EB.

It is not surprising that we only detect starspot modulations from the primary, given the steepness of the stellar mass-luminosity relation. The starspot modulations from the more massive companion will dominate the light curve, except in the rare case of very nearly equal mass stars. Only 9\% of EBs in our sample have $\delta_{sec}/\delta_{pri} > 0.9$, where we would most likely expect to detect both stars. Furthermore, the slightly subsynchronous population of EBs is more pronounced among EBs with low mass ratios (middle panel of Figure \ref{fig:sync_vs_depth}). If the subsynchronous population is due to differential rotation, as we argue below, then the low mass ratio further supports that the signal originates from the primary star only.  

\subsection{Subsynchronous Rotation}
\label{subsec:subsynchronous}

\begin{figure}[ht]
\centering
\includegraphics[width=3in, trim = 0cm 0cm 0cm 0cm, clip]{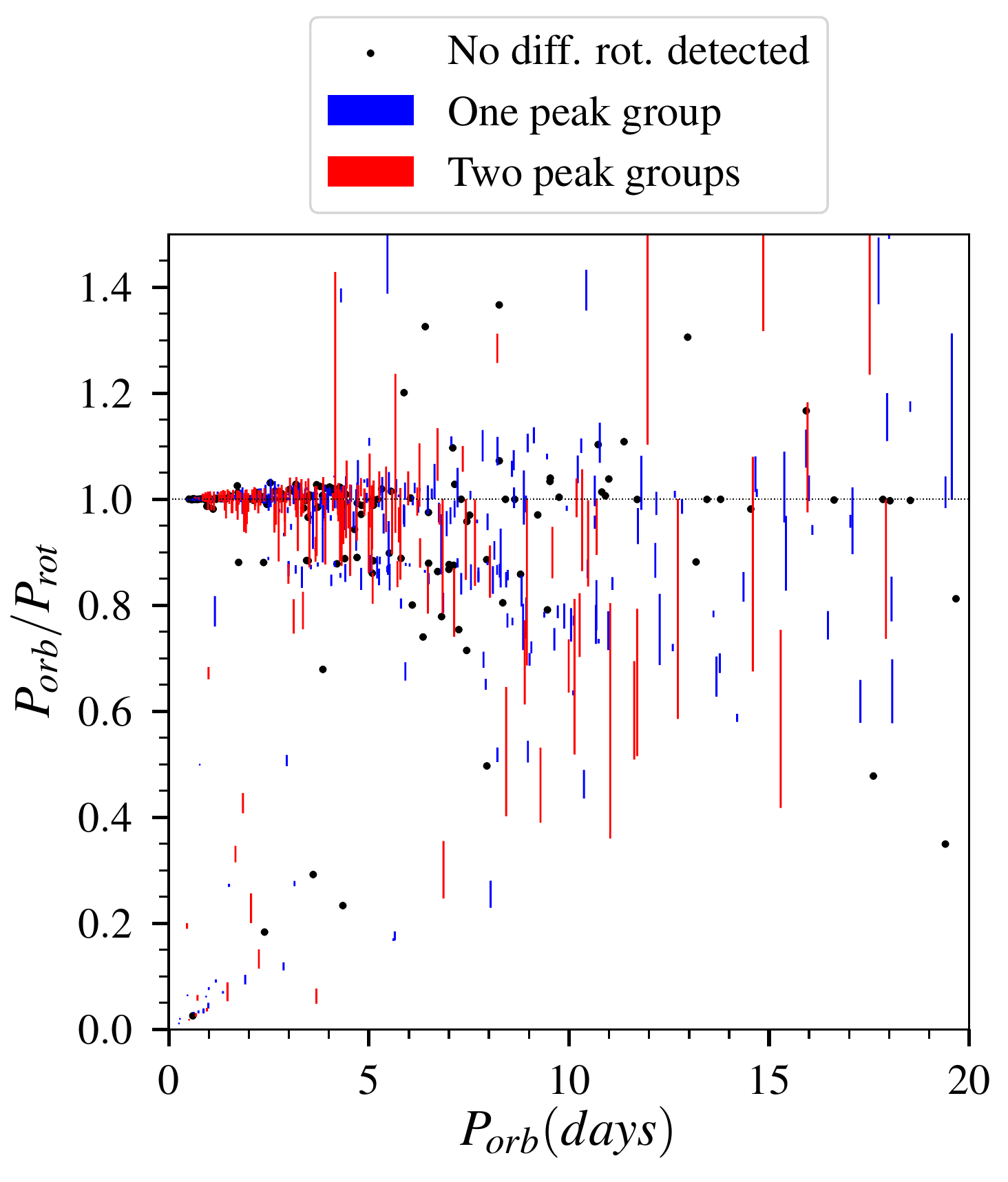}
\caption{The distribution of period ratio versus orbital period. Vertical lines indicate EBs with multiple rotation periods \change{due to differential rotation}. Red lines are for EBs with two periodogram peak groups, and blue lines are for one peak group. Black circles indicate EBs with only one rotation period measurement\change{, for which differential rotation was not detected.}}
\label{fig:multi p_orb/p_rot}
\end{figure}

Given the above results, we will assume that we are detecting differential rotation on the primary star, and now demonstrate how differential rotation explains the subsynchronous population of EBs. \change{To help illustrate this}, Figure \ref{fig:multi p_orb/p_rot} shows \change{the period ratio diagram, with the range of rotation periods due to differential rotation indicated by vertical lines.}

Below 10 days, the EBs are synchronized to the rotation period at the equator. As the orbital period increases, so does the rotation period. As is shown in Figure \ref{fig:shear vs p_rot}, there is a larger amount of relative shear at longer rotation periods. Because of this, the measured values of the period ratio decrease with orbital period. We can then envision an envelope in the $P_{orb}/P_{rot}$-$P_{orb}$ space that stars can occupy. It extends from the synchronization line (or slightly above), and expands downwards. The lower edge of the envelope is dictated by the maximum amount of relative shear possible at a given rotation period, which appears to be roughly 15 to 20\%. Stars could then lie anywhere in this envelope depending on the distribution of their starspots.

% Figure 13: P_orb/P_rot ranges vs P_orb

In this scenario, EBs with no detected differential rotation (black circles \change{in Figure \ref{fig:multi p_orb/p_rot}}) have primaries with starspots that exist only in a narrow latitude range. However, the spots could occur at any latitude, which explains why the black points are distributed throughout the envelope. The single periodogram peak group category (blue vertical lines) have spots in a relatively narrow latitude range, but some differential rotation is detected within this latitude range. In contrast, the two spot group category (red vertical lines) have spots at a large latitude range. In the extreme case, there are spots near the equator and near the poles, so that the vertical line spans the entire envelope. The latitude distribution of spots may also vary over time due to activity cycles.

The subsynchronous population does not extend below orbital periods of approximately two days. There may be very little differential rotation on the most rapidly rotating, tidally synchronized stars. In that case, the higher latitude starspots will have the same rotation period as the equator. Alternatively, the starspots could preferentially be located near the equator in these rapidly rotating stars.

The differential rotation scenario is consistent with two expectations from previous studies: that rapidly rotating stars have less differential rotation than the Sun \citep{CollierCameron2007,Kuker2011}; and that rapidly rotating stars have starspots near their poles \citep{Strassmeier2002}. On the Sun, latitudes $\pm50^{\circ}$ rotate roughly 13\% slower than the equator \citep{Beck2000}, whereas the maximum latitude where sunspots occur is roughly $30^{\circ}$. This implies that the subsynchronous EBs have starspots at higher latitudes than the Sun, perhaps near the poles. If the subsynchronous starspot modulation originates from the poles, then the total equator to pole relative shear is $\alpha \approx 0.13$, compared to approximately 0.3 on the Sun. This is consistent with less differential rotation than the Sun. 

A combination of ellipsoidal variations and starspot modulations is an alternate explanation for the EBs with two periodogram peaks. In this case, the ellipsoidal variations cause the peak at the orbital period, and the starspot modulations cause the subsynchronous peak. However, when we folded the light curves at the orbital period, they showed no evidence of the ellipsoidal variations. We conclude that the periodicity is originating from starspot modulations.

Throughout this discussion, we have assumed that the stars are tidally synchronized to the rotation period at the equator. It is possible that the subsynchronous EBs have achieved resonance locking with convective motions or gravity modes, rather than the surface rotation \citep{Burkart2014}. Alternatively, the EBs could be synchronized to the rotation rate of the radiative core if the tidal energy is dissipated there \citep{Witte2002}. In any case, these results provide a new and important test for tidal theory.

\section{Additional Results}
\label{sec:additional results}

\subsection{Starspot Occultations on a Candidate RS CVn System}
\label{subsec:evolved stars}

We briefly highlight the interesting EB KIC 10614158. It is listed in the KEBC as having an orbital period of 4.46 days, and only primary eclipses. It has an effective temperature of 4600 K according to the \Kepler Input Catalog. Visual inspection of the light curve shows that every other eclipse has a completely flat bottom, while the intervening eclipses have bumps that appear to be spot occultations\footnote{See \citet{Silva2003} and B. Morris et al. (submitted) for examples of spot occultations by planets orbiting main sequence stars.}. Some flares and instrument-related discontinuities are also visible.

% Figure 14: Candidate RS CVn
\begin{figure}[ht]
\centering
\includegraphics[width=3in, trim = 0cm 0cm 0cm 0cm, clip]{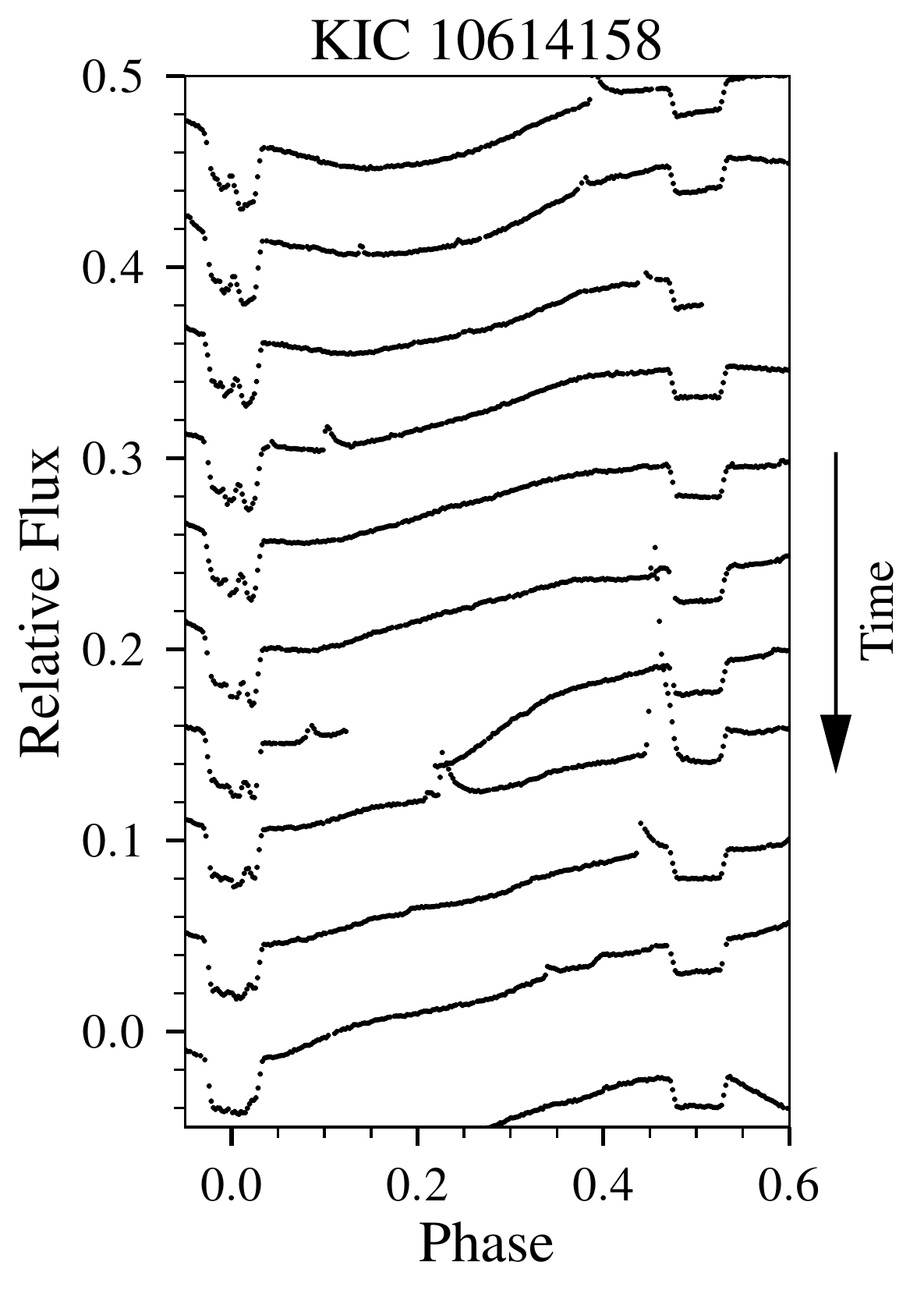}
\caption{Successive eclipses for the candidate RS CVn system KIC 10614158. Time increases towards the bottom of the figure. Spot occultations are visible in the primary eclipses near phase zero, while the secondary eclipses at phase 0.5 have flat bottoms.}
\label{fig:rs cvn}
\end{figure}

This pattern is demonstrated in Figure \ref{fig:rs cvn}, where each successive eclipse is vertically offset for clarity. The spot occultations occur near phase zero and move in phase over time. This pattern is inconsistent with only primary eclipses. We instead argue that KIC 10614158 has an orbital period of 8.92 days. The primary eclipses with spot occultations occur when the main-sequence star passes in front of the larger, more luminous evolved star, which has spots. The secondary eclipses occur when the main-sequence star disappears behind the evolved star. KIC 10614158 is a good target for further investigation, as it provides a unique opportunity to study the tidal interaction and starspot distribution of evolved stars.

\subsection{The Period-Color Relation for Contact Binaries}
\label{subsec:contact variables}

\change{There is a well known relation between the orbital periods and photometric colors of contact binaries (e.g, \citealt{Eggen1967,Rucinski1994,Rubenstein2001}). These stars are filling their Roche lobes, directly linking the orbital period to the} stellar radius, mass, and photometric color. The redder contact binaries \change{(larger color indices)} have shorter orbital periods, implying smaller stellar radii.

\begin{figure}[ht]
\centering
\includegraphics[width=3in, trim = 0cm 0cm 0cm 0cm, clip]{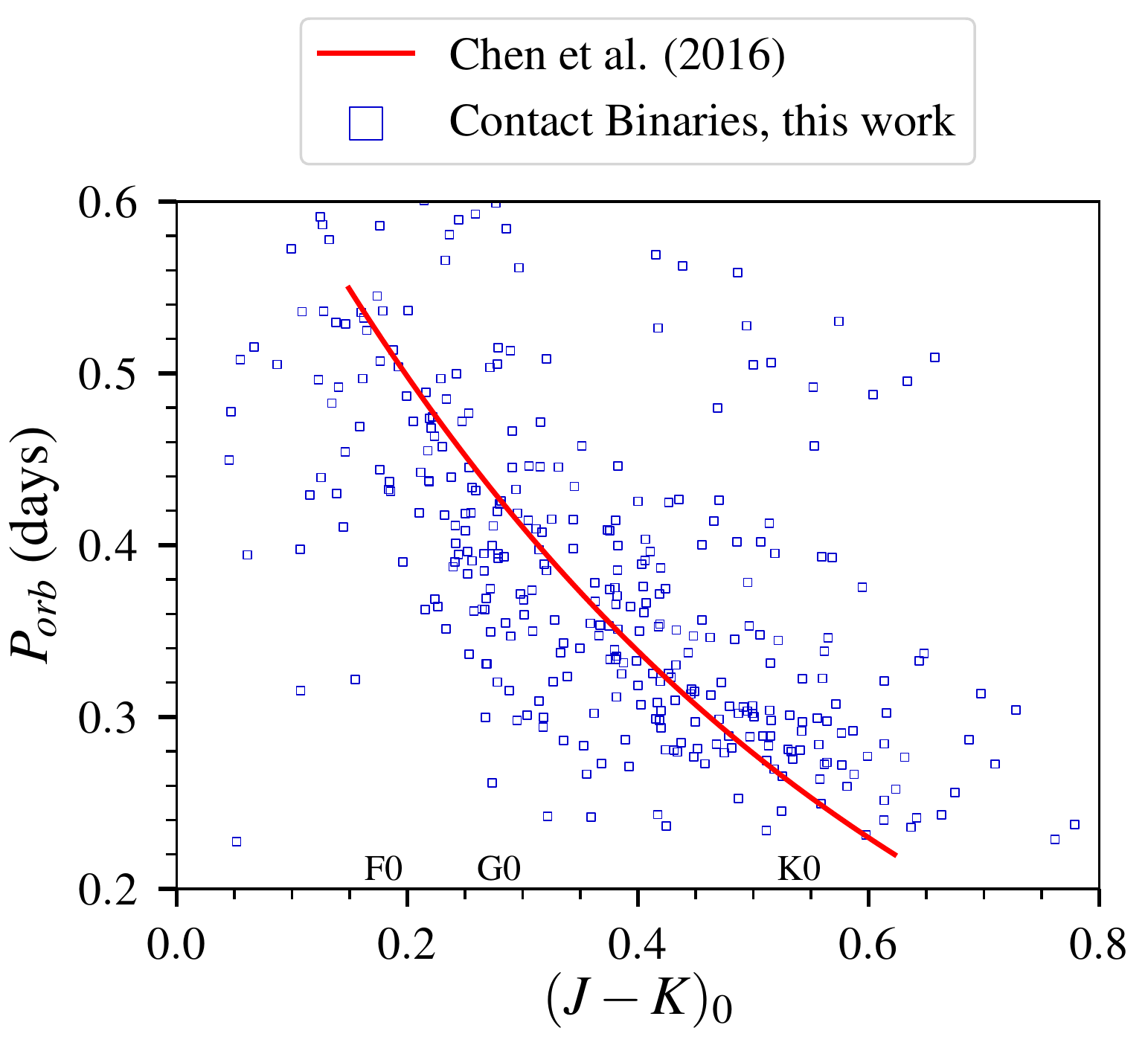}
\caption{The distribution of orbital period versus \change{dereddened $J-K$} color for contact binaries. \change{The red curve shows the empirical period-color relation from \citet{Chen2016}. Spectral types from \citet{Covey2007} are shown for reference.}}
\label{fig:contact binaries}
\end{figure}

Figure \ref{fig:contact binaries} shows the distribution of orbital periods versus \change{dereddened $J-K$} colors for EBs with ellipsoidal variations and orbital periods less than 0.6 days. These EBs appear to be contact binaries based upon their light curves. \change{For comparison, we show the period-color relation from \citet{Chen2016}, based on a fit to over 6000 contact binaries collected from the literature. Their relation is a good fit to our sample as well.}

% Figure 15: Contact binaries

\section{Conclusion}
\label{sec:summary}

We have analyzed 2278 EBs in the \Kepler Eclipsing Binary Catalog for evidence of tidal synchronization. EBs were visually classified based on their out-of-eclipse variability as having starspot modulations, ellipsoidal variations, pulsations, other out-of-eclipse variability, or no out-of-eclipse periodic variability. For EBs with starspot modulations, we measured multiple rotation periods using a combination of the autocorrelation function and the Lomb-Scargle periodogram. Our main results are summarized as follows:

\begin{itemize}

\item At orbital periods less than 10 days, most EBs are tidally synchronized. Below two days, 94\% of EBs are synchronized, defined as having rotation periods within 10\% of their orbital periods. At orbital periods between two and ten days, this number is 72\%.

\item There is a population of subsynchronous EBs, which has not been observed in previous studies. Between orbital periods of two and ten days, 15\% of EBs have rotation periods that are typically 13\% longer than their orbital periods. 

\item This subsynchronous population has low eccentricities, slightly favors lower mass ratios, and shows no strong correlation with mass for \change{F,} G and K type primaries.

\item We demonstrated that the subsynchronous population is consistent with differential rotation. Over three quarters (77\%) of EBs with starspot modulations have multiple rotation periods, which are likely originating from differentially rotating active latitudes on the primary star. The primaries are likely synchronized to the rotation period at the equator, and spots near the poles cause the measured rotation period to be longer than the orbital period. Some EBs appear have spots near both the equator and poles, perhaps due to activity cycles or a range of differential rotation profiles.

\item At an orbital period of roughly 10 days, there is a transition from primarily circularized and synchronized EBs to primarily eccentric and pseudosynchronized EBs. This transition is in good agreement with the predicted and observed tidal circularization period for Milky Way field binaries.

\item Our rotation period catalog mostly contains EBs with \change{F,} G and K type primary stars, because the \Kepler target selection favors solar-type stars, and because starspot modulations are not found on earlier type stars. This is beneficial in that it greatly increases the number of published rotation period measurements for such binaries. There is no clear difference in synchronization between \change{F,} G and K primaries, suggesting that primary mass is not an important factor in synchronization over the relatively small mass range of \change{F,} G and K stars.

\item For both small and nearly equal mass ratios, EBs with periods less than 10 days are highly synchronized. Beyond ten days, EBs with small mass ratios are somewhat less synchronized than EBs with nearly equal mass ratios. 

\end{itemize}

The tidal interaction of close binaries is an important aspect of stellar astrophysics, but also has much broader implications for stellar populations. Our results represent a substantial increase in the observational data for tidally interacting late-type binaries, and offer many opportunities for further investigation. The transition from circular, synchronized EBs to eccentric, pseudosynchronized EBs is worthy of additional modeling to better understand the complex dynamics at work. The same can be said for the differential rotation mechanism we introduced to explain the population of subsynchronous EBs. 

We are currently expanding our analysis to the \textit{K2} mission, which has observed binaries with a wider range of spectral types and ages. Combined with improved stellar parameters from the \textit{Gaia} mission, we will be able to consider interaction in a Galactic context.

\acknowledgements
This work was supported by NSF grant AST13-12453, the University of Washington College of Arts and Sciences, the Washington Research Foundation, and the University of Washington Provost's Initiative for Data-Intensive Discovery.

The authors are grateful to Brett Morris, Leslie Hebb, and Rory Barnes for helpful suggestions during the preparation of this paper.  

This work includes data collected by the Kepler mission. Funding for the Kepler mission is provided by the NASA Science Mission Directorate. This work has made use of NASA's Astrophysics Data System Bibliographic Services.

\software{gatspy \citep{VanderPlas2015,VanderPlas2016}, h5py (Andrew Collette and contributors, 2008, \url{http://h5py.alfven.org}), IPython \citep{Perez2007}, kplr (Daniel Foreman-Mackey, \url{http://dan.iel.fm/kplr}), Matplotlib \citep{Hunter2007}, NumPy \citep{vanderWalt2011}, Pandas \citep{McKinney2010}, SciPy \citep{Jones2001}}

\appendix
\section{Asynchronous EBs with $P_{orb} < 10$ Days}

% Outliers table
\startlongtable
\begin{deluxetable}{l}
\label{tab:outliers}

\tabletypesize{\small}
\tablewidth{0pt}
\tablecaption{Asynchronous Short Period Systems}

\tablehead{\colhead{Kepler ID}}

\startdata
\hline  
\multicolumn{1}{c}{KEBC Orbital Period Corrected}\\
\hline  
1161345 ($P_{orb} = 2P_{KEBC}$) \\
2558370 ($P_{orb} = 2P_{KEBC}$) \\
4454219 ($P_{orb} = 2P_{KEBC}$) \\
4912991 ($P_{orb} = 0.5P_{KEBC}$) \\
8409588 ($P_{orb} = 2P_{KEBC}$) \\
9084778 ($P_{orb} = 2P_{KEBC}$) \\
9592575 ($P_{orb} = 2P_{KEBC}$) \\
9597411 ($P_{orb} = 2P_{KEBC}$) \\
10614158 ($P_{orb} = 2P_{KEBC}$) \\
10848459 ($P_{orb} = 2P_{KEBC}$) \\
11303811 ($P_{orb} = 0.5P_{KEBC}$) \\
\hline  
\multicolumn{1}{c}{Possible false positives}\\
\hline  
4929299 \\
5642620 \\
6370120 \\
7051984 \\
8176653 \\
9478836 \\
9642018 \\
10338279 (SB2 in \citealt{Kolbl2015}) \\
10407221 \\
10857519 (SB2 in \citealt{Kolbl2015}) \\
12170648 \\
12255382 \\
\hline  
\multicolumn{1}{c}{Possible planets or brown dwarfs}\\
\hline  
3970233 \\
5369827 \\
7269493 \\
7376983 \\
7763269 (SB2 in \citealt{Kolbl2015})\\
9752973 (SB2 in \citealt{Kolbl2015}) \\
9880467 \\
9895004 \\
10395543 \\
10925104 \\
\hline  
\multicolumn{1}{c}{Asynchronous short period EBs}\\
\hline  
2445975 \\
3443790 \\
3459199 \\
3848972 \\
4367544 \\
4456622 \\
4946584 \\
5372966 \\
5648449 \\
6956014 \\
7684873 \\
7838906 \\
8906676 \\
8938628 \\
9266285 \\
9579499 \\
10613718 \\
11147460 \\
11231334 \\
11548140 \\
11560037 \\
\hline  
\multicolumn{1}{c}{Pseudosynchronized EBs}\\
\hline  
5024292 \\
5025294 \\
7376500 \\
9971475 \\
10287248 \\
10923260 \\
12470530\\
\enddata

%\tablecomments{}
%\tablenotetext{}{}

\end{deluxetable}

\newpage

{\it Facilities: \facility{Kepler}}

\bibliography{tides_paper}

\end{document}